\documentclass{aa}
\usepackage{times}
\usepackage{graphics}
\usepackage{epsf}
\usepackage{pifont}
\begin{document}
\topmargin0.0cm
%
%\thesaurus{06(03.20.1; 08.03.4; 08.09.2: NML\,Cyg; 08.13.2; 08.19.3; 13.09.6)}
%
\title{
Bispectrum speckle interferometry observations and radiative transfer modelling
of the red supergiant \object{NML\,Cyg}}

\subtitle{Multiple dust-shell structures evidencing previous superwind phases}
%
%%\thanks{Based on observations with  the SAO 6\,m telescope operated by the
%%Special Astrophysical Observatory, Russia}
%
\author{
T.\ Bl\"ocker\inst{1}\and 
Y.\ Balega\inst{2}\and
K.-H.\ Hofmann\inst{1}\and
G. Weigelt\inst{1}
}
\institute{
Max--Planck--Institut f\"ur Radioastronomie, Auf dem H\"ugel 69,
D--53121 Bonn, Germany \\
(bloecker@mpifr-bonn.mpg.de, hofmann@mpifr-bonn.mpg.de,
 weigelt@mpifr-bonn.mpg.de)
\and
Special Astrophysical Observatory, Nizhnij Arkhyz, Zelenchuk region,
Karachai--Cherkesia, 35147, Russia (balega@sao.ru)
}
\offprints{T.\,Bl\"ocker}
%%%\offprints{\hbox{T.\,Bl\"ocker\,(bloecker@mpifr-bonn.mpg.de)}}
%%%\mail{speckle@mpifr-bonn.mpg.de}
%

%%%\date{revised version: \today}
%%%% \date{revised version: \today}
%%%\date{received ~~~~ / accepted ~~~~}
\date{Received date /  accepted date}
\titlerunning{
Bispectrum speckle interferometry observations and radiative transfer models
of \object{NML\,Cyg}}
\authorrunning{T.\ Bl\"ocker et al.}
\abstract{
\object{NML\,Cyg} is a highly evolved OH/IR supergiant,
one of the most prominent
infrared objects due to its strong obscuration by dust, and supposed to be
among the most luminous supergiants in the galaxy.
We present the first diffraction-limited
$2.13\,\mu$m observations of \object{NML\,Cyg} with
73\,mas resolution. The speckle interferograms were obtained with the 6\,m
telescope at the Special Astrophysical Observatory, and the
image reconstruction is based on the bispectrum speckle-interferometry method.
The visibility function declines towards the diffraction limit to $\sim 0.6$.\\
Radiative transfer calculations have been carried out to model the spectral
energy distribution, given by ground-based photo\-metry and ISO spectroscopy,
and our $2.13\,\mu$m visibility function. Additionally,
mid-infrared visibility functions at $11\,\mu$m were considered.
The observed dust shell properties do not appear to be in accordance with
standard single-shell (uniform outflow) models but seem to require
multiple components. Considering previous periods of
enhanced mass-loss, various density enhancements 
in the dust shell were taken
into account. An extensive grid of models was calculated for different
locations and strenghts of such superwind regions in the dust shell.
To match the observations from the optical to the
sub-mm domain requires at least two superwind regions embedded in the shell.
The best model includes a dust shell with a temperature of 1000\,K at its
inner radius of $6.2 R_{\ast}$, a close embedded superwind shell extending
from  $15.5 R_{\ast}$ to  $21.7 R_{\ast}$
with an amplitude (factor of density enhancement) of 10,
and a far-out density enhancement at  $186 R_{\ast}$ with an amplitude of 5.
The angular diameters of the central star and of the inner rim
of the dust shell amount to 16.2\,mas and 105\,mas, resp.
The diameter of the embedded close superwind region extends from
263\,mas to 368\,mas, and the inner boundary of the distant superwind region
has a diameter of 3\farcs 15.
In the near-infrared the dust condensation zone is limb-brightened leading
to a corresponding ring-like intensity distribution.
The grain sizes, $a$, were found to be in accordance with
a standard distribution function,
$n(a)$\,$\sim$\,$a^{-3.5}$, with $a$ ranging between
$a_{\rm min}$\,=\,0.005\,$\mu$m and $a_{\rm max}$\,=\,0.15\,$\mu$m.
The bolometric flux  amounts to
$F_{\rm bol} = 3.63 \cdot 10^{-9}$\,Wm$^{-2}$
corresponding to a central-star 
luminosity of $L/L_{\odot} = 1.13 \cdot 10^{5} \cdot (d/{\rm kpc})^{2}$.
Within the various parts of the dust shell, $1/r^{2}$ density distributions
could be maintained differing only in their amplitude $A$.
A slight improvement of the far-infrared properties can be obtained if
a shallower density distribution of $\rho \sim 1/r^{1.7}$ is considered
in the distant superwind region.
The present-day mass-loss rate was determined to be
$\dot{M} = 1.2 \cdot 10^{-4}$\,M$_{\odot}$/yr.
The inner embedded superwind shell corresponds to a phase
of enhanced mass-loss (with amplitude 10)
in the immediate history of \object{NML\,Cyg} which began 59\,yr ago and
lasted for $\sim 18$\,yr.
Correspondingly, the outer superwind region is due to
to a high mass-loss period (amplitude 5) which terminated 529\,yr ago.
\keywords{
Techniques: image processing --
Circumstellar matter --
Stars: individual: NML\,Cyg --
Stars: mass--loss --
Stars: supergiants --
Infrared: stars}
}
\maketitle
\section{Introduction} \label{Sintro}
The star \object{NML\,Cyg}
(= \object{V\,1489~Cyg} = \object{IRC\,+40\,448}) belongs to the most
prominent infrared objects of the northern hemisphere. It was
discovered by Neugebauer, Martz \& Leighton (\cite{NML65})
being an extremely red object in their infrared survey.
This redness is due to \object{NML\,Cyg}'s strong obscuration by
circumstellar dust (e.g.\ Herbig \& Zappala \cite{HerZap70},
Hyland et al.\ \cite{HylEtal72}).
The spectral type has been determined to be M6\,III
(Herbig \& Zappala \cite{HerZap70}), however also luminosity classes of up to
Ia have been derived (e.g.\ Johnson \cite{John68},
                         Low et al.\ \cite{LowEtal70}).

Various distance estimates exist for \object{NML\,Cyg} ranging from
200\,pc (Herbig \& Zappala \cite{HerZap70}) up to 3400\,pc
(Rowan-Robinson \& Harris \cite{RowRobHar83}).
The most likely distance appears to be $\sim 2$\,kpc due to the
immediate proximity to the  Cyg OB2 association
(Morris \& Jura \cite{MorJur83}).
\object{NML\,Cyg} is partially surrounded by an H\,II region
(Goss et al.\ \cite{GossEtal74}, Gregory \& Seaquist \cite{GreSea76},
 Habing et al.\ \cite{HabEtal82}) which is supposed to be the result of
ionizing radiation from the Cyg OB2 association
(Morris \& Jura \cite{MorJur83}).

\object{NML\,Cyg} is associated with
OH  (Wilson et al.\ \cite{WilEtal70}, Masheder et al.\ \cite{MashEtal74},
     Benson \& Mutel \cite{BenMut79}, Bowers et al.\ \cite {BowEtal83}),    
SiO (Snyder \& Buhl \cite{SnyBuhl75}, Dickinson et al.\ \cite{DickEtal78},
     Morris et al.\ \cite{MorEtal79})
and H$_{2}$O (Schwartz \& Barrett \cite{SchwBar70},
              Schwartz et al.\  \cite{SchwEtal74},
              Richards et al.\  \cite{RichEtal96} )
maser sources.
The outflow velocity for the circumstellar shell was measured to be
21\,km/s in the CO(2-1) line (Knapp et al.\ \cite{KnaEtal82}),
24\,km/s in H$_{2}$O emission profiles at 22\,GHz
(Engels et al.\ \cite{EngEtal86}),
26\,km/s in thermal SiO emission at 87\,GHz
(Lucas et al.\ \cite{LucEtal92}), 
and 28\,km/s in OH emission profiles at 1612\,MHz
(Bowers et al.\ \cite{BowEtal83}). Mass-loss rates,
based on the CO and OH measurements and 
scaled to a distance of 1.8\,kpc, 
are
$1.5 \cdot 10^{-4}$\,M$_{\odot}$/yr and $1.6 \cdot 10^{-4}$\,M$_{\odot}$/yr,
resp.

Thus \object{NML\,Cyg} can be believed to be a highly evolved OH/IR supergiant
of very large luminosity which suffers from strong mass-loss and is
highly enshrouded by dust. Morris \& Jura (\cite{MorJur83}) estimated its
luminosity to be $5 \cdot 10^{5}$\,L$_{\odot}$ at the distance of 2\,kpc
(correspondig to an initial mass of $\sim 40$\,M$_{\odot}$). Consequently
\object{NML\,Cyg} appears to be among the most luminous supergiants in the
Galaxy (Richards et al.\ \cite{RichEtal96}).

There are indications that the dust shell of \object{NML\,Cyg} may possibly
deviate from spherical symmetry. Strong polarization was found by, e.g.,
Forbes (\cite{For67}),
Kruzewski (\cite{Kru71}), and Dyck et al.\ (\cite{DyckEtal71}). 
The dust shell morphology appears to be complex. For instance, 
Diamond et al.\ (\cite{DiaEtal84}) found the OH envelope to be asymmetric,
and Richards et al.\ (\cite{RichEtal96}) concluded the existence of a bipolar
outflow from interferometric observations of the
H$_{2}$O maser.

Previous infrared speckle interferometric observations have been reported by
Sibille et al.\ (\cite{SibEtal79}), Dyck et al.\ (\cite{DyckEtal84}),
Ridgway et al.\ (\cite{RidgEtal86}), 
Fix \& Cobb (\cite{FixCob88}) and Dyck \& Benson (\cite{DyckBen92}).
Recent results from mid-infrared interferometry are given in
Monnier et al.\ (\cite{MonEtal97}), Danchi et al.\ (\cite{DanEtal99}), 
and Sudol et al.\ (\cite{SudEtal99}).

In this paper we present diffraction-limited  73\,mas
bispectrum speckle-interferometry  observations
of the dust shell of \object{NML\,Cyg}.
Radiative transfer calculations have been performed to model the spectral
energy distribution as well as visibilities at different wavelengths.
\section{\hbox{Near-infrared observations and data reduction}}
The \object{NML\,Cyg} speckle interferograms
were obtained with the Russian
6\,m telescope at the Special Astrophysical Observatory
on June 13 and 14, 1998.
The speckle data were recorded
with our NICMOS-3 speckle camera
(HgCdTe array, 256$^2$ pixels,
sensitivity from 1 to 2.5\,$\mu$m,
frame rate 2 frames/s)
through a continuum filter with a
centre wavelength of 2.13\,$\mu$m and a bandwidth of 0.02\,$\mu$m.
Speckle interferograms of the unresolved star \object{HIP 102098}
were taken for the
compensation of the speckle interferometry transfer function.
The observational
parameters were as follows:
exposure time/frame 25~ms; number of frames 5000
(2000 of \object{NML\,Cyg} and 3000 of \object{HIP 102098});
2.13\,$\mu$m seeing (FWHM) $\sim$1\farcs0;
field of view 7\farcs8$\times$7\farcs8; pixel size 30.52\,mas.
A diffraction-limited image of  \object{NML\,Cyg}
with 73\,mas resolution was reconstructed from
the speckle interferograms using the bispectrum speckle-interferometry method
(Weigelt \cite{Wei77}, Lohmann et al.\ \cite{LohWeiWir83},
Hofmann \& Weigelt \cite{HofWei86}, Weigelt \cite{Wei91}).
The bispectrum of each frame consisted of $\sim$37 million elements.
The modulus of the object Fourier transform (visibility) was
determined  with the speckle interferometry method (Labeyrie \cite{Lab70}).

%%%%%%%%%%%%%%%%%%%%%%%%%%%%%%%%%%%%%%%%%%%%%%%%%%%%%%%%%%%%%%%%%%%%%%%%%%%%%
%%%% 2D Powerspektrum + visibility nebeneinander
%%%%%%%%%%%%%%%%%%%%%%%%%%%%%%%%%%%%%%%%%%%%%%%%%%%%%%%%%%%%%%%%%%%%%%%%%%%%%
\begin{figure}
\parbox{4.0cm}{
\epsfxsize=3.7cm
\vspace*{-2mm}
%%% \epsfbox[-30 -30 226 226]{./plots/objpow.nord_14-40+44_p2.5.ima.newlut.eps}}
\epsfbox[-30 -30 226 226]{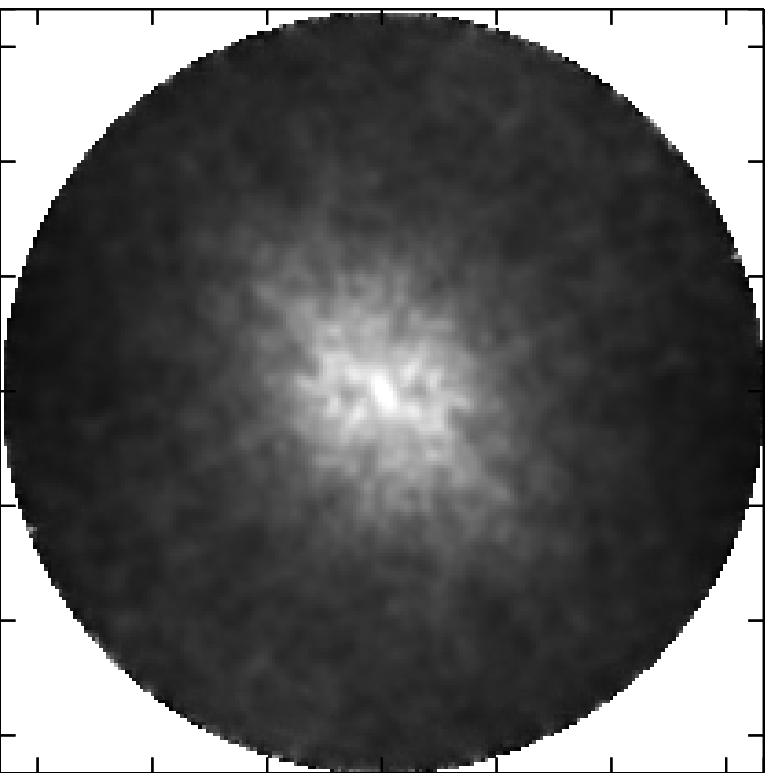}}
\epsfxsize=4.2cm
\parbox{4.8cm}{
\hspace*{2mm}
%%% \epsfbox[95 50 320 264]{./plots/objvis2.radarcs_4.vis.eps}}
\epsfbox[95 50 320 264]{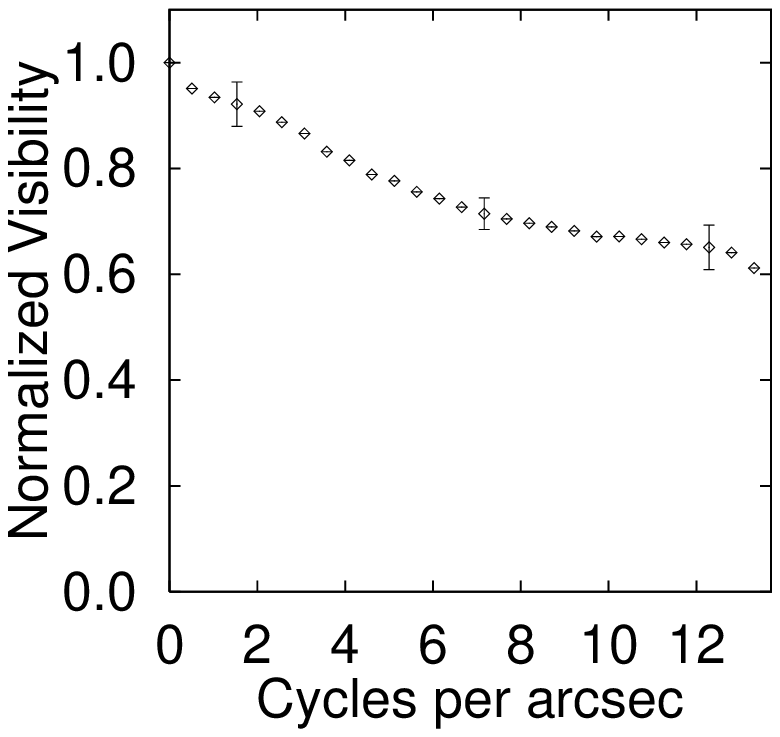}}
\caption{{\bf Left:} Two-dimensional 2.13\,$\mu$m visibility function of
\object{NML\,Cyg} shown up to the diffraction limit (see right panel).
The bright central structure shows that the central object is
surrounded by a resolved dust shell.
{\bf Right:}
Azimuthally averaged  2.13\,$\mu$m visibility of \object{NML\,Cyg} with
error bars for selected frequencies.
}
\label{Fvisi} 
\end{figure}
%%%%%%%%%%%%%%%%%%%%%%%%%%%%%%%%%%%%%%%%%%%%%%%%%%%%%%%%%%%%%%%%%%%%%%%%%%%%%
%%% Contour plot
%%%%%%%%%%%%%%%%%%%%%%%%%%%%%%%%%%%%%%%%%%%%%%%%%%%%%%%%%%%%%%%%%%%%%%%%%%%%%
\begin{figure}
\epsfxsize=4.8cm
\parbox{5.5cm}{
\hspace*{0.5cm}
%%%\epsfbox{./plots/NMLCyg-K.ima.5e.eps}}
\epsfbox{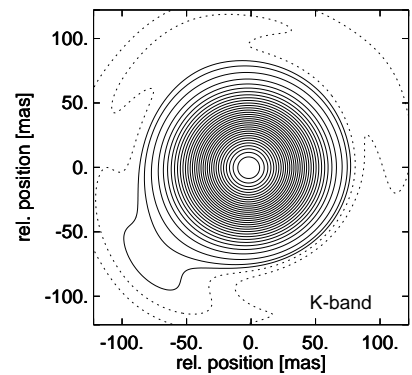}}
\parbox{3.2cm}{
\caption{
Contour plot of the bispectrum speckle interferometry $2.13 \mu$m-image
of \object{NML\,Cyg}. Contour levels are plotted for 2\%, 3\%, 4\%, and 5\%
of peak intensity, 
and from 7\% to 97\% of peak intensity in steps of 3\%. The dashed contours
(2\%, 3\%) refer to the noise floor. North is up and east is to the left.
} \label{Fcont}
}
\end{figure}
%%%%%%%%%%%%%%%%%%%%%%%%%%%%%%%%%%%%%%%%%%%%%%%%%%%%%%%%%%%%%%%%%%%%%%%%%%%%%
%%% Radial plots
%%%%%%%%%%%%%%%%%%%%%%%%%%%%%%%%%%%%%%%%%%%%%%%%%%%%%%%%%%%%%%%%%%%%%%%%%%%%%
\begin{figure}
\epsfxsize=5.5cm
\parbox{5.5cm}{
%%% \epsfbox[42 13 397 325]{./plots/f_NMLCyg.azimut.eps}}
\epsfbox[42 13 397 325]{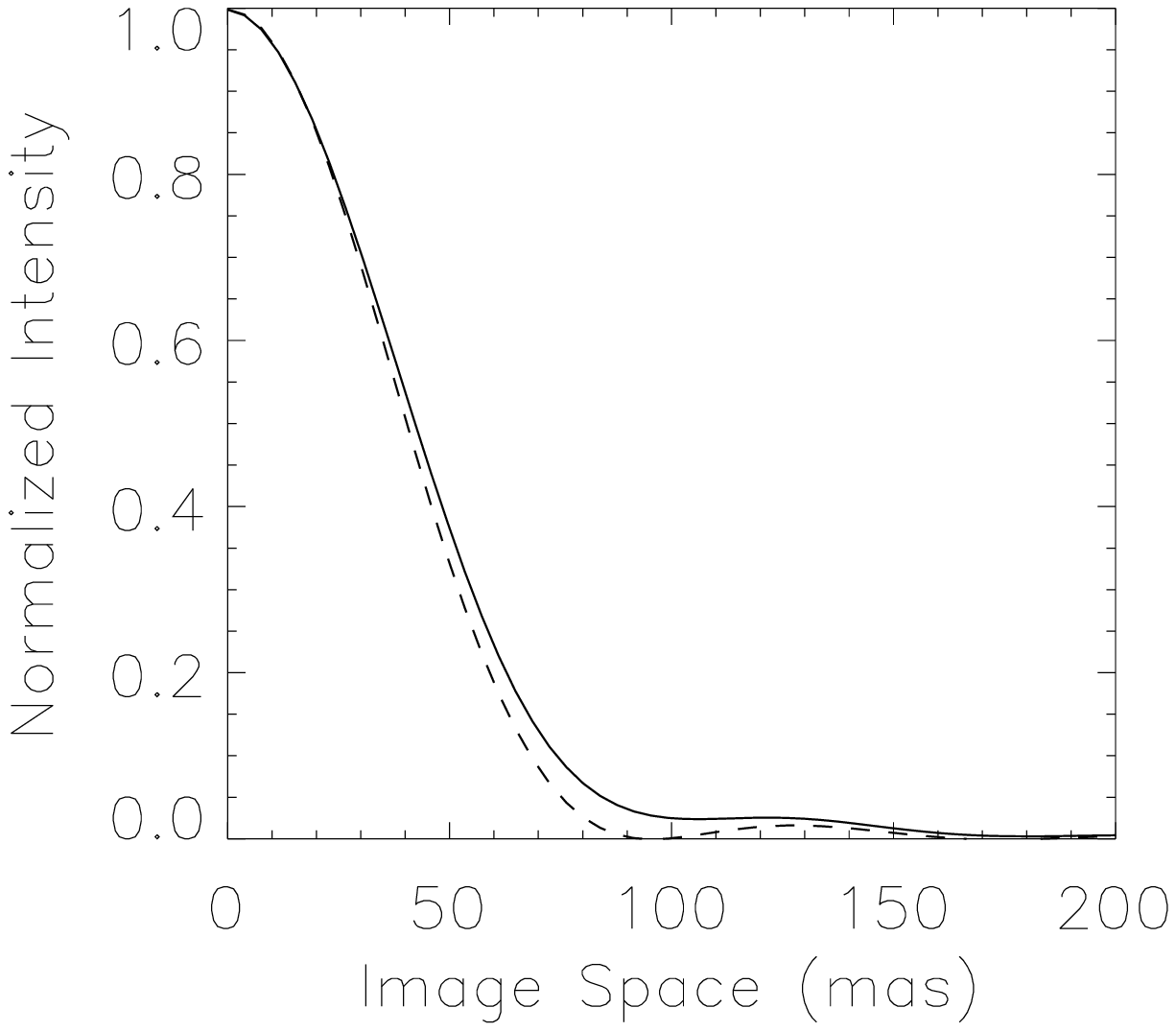}}
\parbox{3.2cm}{
\caption{
Azimuthally averaged radial plots of the reconstructed diffraction-limited
$2.13 \mu$m-images of
\object{NML\,Cyg} (solid line) and of the unresolved star 
\object{HIP\,102098} (dashed line).
} \label{Fradial}
}
\end{figure}
%%%%%%%%%%%%%%%%%%%%%%%%%%%%%%%%%%%%%%%%%%%%%%%%%%%%%%%%%%%%%%%%%%%%%%%%%%%%%
%
Figure~\ref{Fvisi}
shows the reconstructed 2.13\,$\mu$m visibility function of
\object{NML\,Cyg}, whereas Fig.~\ref{Fcont} presents a contour plot of
the reconstructed image.
There is only little
evidence for deviations from spherical symmetry. 
The object appears elongated at the 4\% intensity level,
which, however, is already close to the underlying noise floor, 
with a position angle of $\sim 140\degr$, 
However, it is noteworthy that this position angle is close to
the one found by $5\,\mu$m interferometric measurements
($\sim 140\degr$-$150\degr$; McCarthy 1979) and to 
those found for the  symmetry axes of the
SiO ($\sim 140\degr$; Boboltz \& Marvel 2000),
H$_{2}$O ($\sim 130\degr$; Richards et al.\ 1996), and
OH ($\sim 150\degr$; Masheder et al.\ 1974) maser distributions.
%%%the one found by Richards et al.\ (1996) for the symmetry axis of the
%%%H$_{2}$O maser distribution ($\sim 150\degr$).
The Gau\ss\ fit FWHM diameter of the dust shell was determined
to be 121\,mas.
Fig.~\ref{Fradial} displays the azimuthally averaged diffraction-limited
images of \object{NML\,Cyg} and the unresolved star HIP 102098.
\section{Mid-infrared visibilities}
\begin{figure}[htbp]
\centering
\epsfxsize=8.8cm
%%%%\mbox{\epsffile{\RHOME/nmlcyg/aa_visi_11_data.ps}}
\mbox{\epsffile{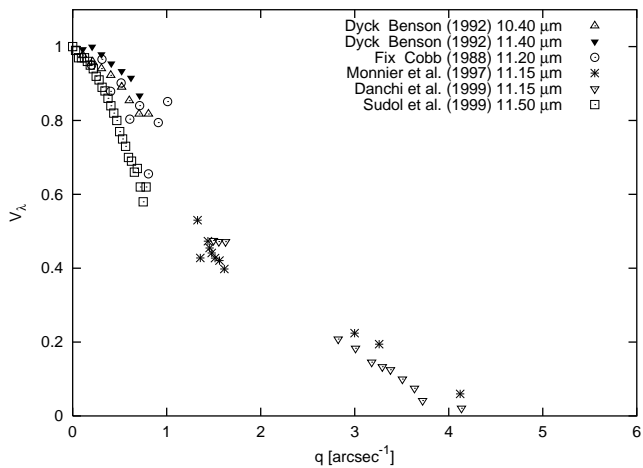}} 
\caption[visi11]
{Mid-infrared visibility functions of \object{NML\,Cyg}
 at $11.2 \mu$m (Fix \& Cobb 1988),
 at  $10.4 \mu$m and $11.4 \mu$m (Dyck \& Benson 1992),
 at  $11.15 \mu$m (Monnier et al.\ 1997; Danchi et al.\ 1999),
 and at $11.5 \mu$m (Sudol et al.\ 1999)}
 \label{Fvisi11}
\end{figure}

Whereas the $K$-band visibility traces the hot dust in the immediate vicinity
of the star, visibility functions in the mid-infrared can
give spatial information on the colder dust located
further out
in the
circumstellar shell where the 9.7\,$\mu$m silicate feature forms. 
In the instance of \object{NML\,Cyg} and its heavy obscuration by dust,
a considerable fraction of the observed flux is radiated at around 9.7\,$\mu$m
and due to the thickness of the shell the silicate feature appears in
absorption (see Fig.~\ref{Fsed}). 
Accordingly complementary $N$-band (11.2\,$\mu$m) interferometry can supply
important constraints for the astrophysical interpretation of the object
by, e.g., dust-shell modelling. 
Fig.~\ref{Fvisi11} refers to corresponding observations in this regime and 
shows the visibility functions at $11.2 \mu$m (Fix \& Cobb 1988),
at  $10.4 \mu$m and $11.4 \mu$m (Dyck \& Benson 1992),
at  $11.15 \mu$m (Monnier et al.\ 1997, Danchi et al.\ 1999),
and at $11.5 \mu$m (Sudol et al.\ 1999). In the following sections
we will focus on the most recent ones, i.e. the highest resolution observation
of Monnier et al.\ (1997) and Danchi et al.\ (1999)
with the ISI heterodyne inter\-fero\-meter in 1993, 1994, 1996, and 1998
and the latest low spatial frequency measurement of
Sudol et al.\ (1999) at the 2.3\,m Wyoming Infrared Observatory in 1997.
%
%%%%%%%%%%%%%%%%%%%%%%%%%%%%%%%%%%%%%%%%%%%%%%%%%%%%%%%%%%%%%%%%%%%%%%%%%%%%%
%%%%%%%%%%%%%%%%%%%%%%%%%%%%%%%%%%%%%%%%%%%%%%%%%%%%%%%%%%%%%%%%%%%%%%%%%%
%
\section{Spectral energy distribution}
Photometric and spectrometric data in the optical and infrared
for \object{NML\,Cyg} are given, e.g., by
Johnson  et al.\ (\cite{JohnEtal65}),
Johnson (\cite{John67}),
Low et al.\ (\cite{LowEtal70}),
Hackwell (\cite{Hack72}),
Dyck et al.\ (\cite{DyckEtal74}),
Strecker \& Ney (\cite{StrNey74}) and
Merrill \& Stein (\cite{MerSt76}).
\object{NML\,Cyg} shows small long-term variations in its spectral energy
distribution (Harvey et al.\ \cite{HarEtal74}, Strecker \cite{Str75}) with
a period of $\sim 1000$\,days. Recently, Monnier et al.\ (\cite{MonEtal97})
presented long-term 10.2\,$\mu$m photometry covering a time basis of more
than 12\,yr. They determined an amplitude of $\sim 0.5$\,mag and a mean period
of $\sim 940$\,days. Thus \object{NML\,Cyg} can be considered as a
semiregular variable.
Monnier et al.\ (\cite{MonEtal97}, \cite{MonEtal98})
investigated the temporal variation of the mid-infrared spectra by means of
8-13\,$\mu$m spectrophotometry from 1991 to 1995 finding an almost constant
spectral shape in this wavelength regime. 
%%%%%% {\tt Monnier 1999 more reddish...? No, IRAS spectra do not fit\\}
Submillimeter (400\,$\mu$m) and millimeter (1.3\,mm) observations are given
by Sopka et al.\ (\cite{SopEtal85}) and Walmsley et al.\ (\cite{WalEtal91}),
resp.
%
%%%%%%%%%%%%%%%%%%%%%%%%%%%%%%%%%%%%%%%%%%%%%%%%%%%%%%%%%%%%%%%%%%%%%%%%%%%%%
\begin{figure}[bthp]
\centering
\epsfxsize=8.8cm
%%%%%%\mbox{\epsffile{\RHOME/nmlcyg/aa_sed_data_ground.ps}}
\mbox{\epsffile{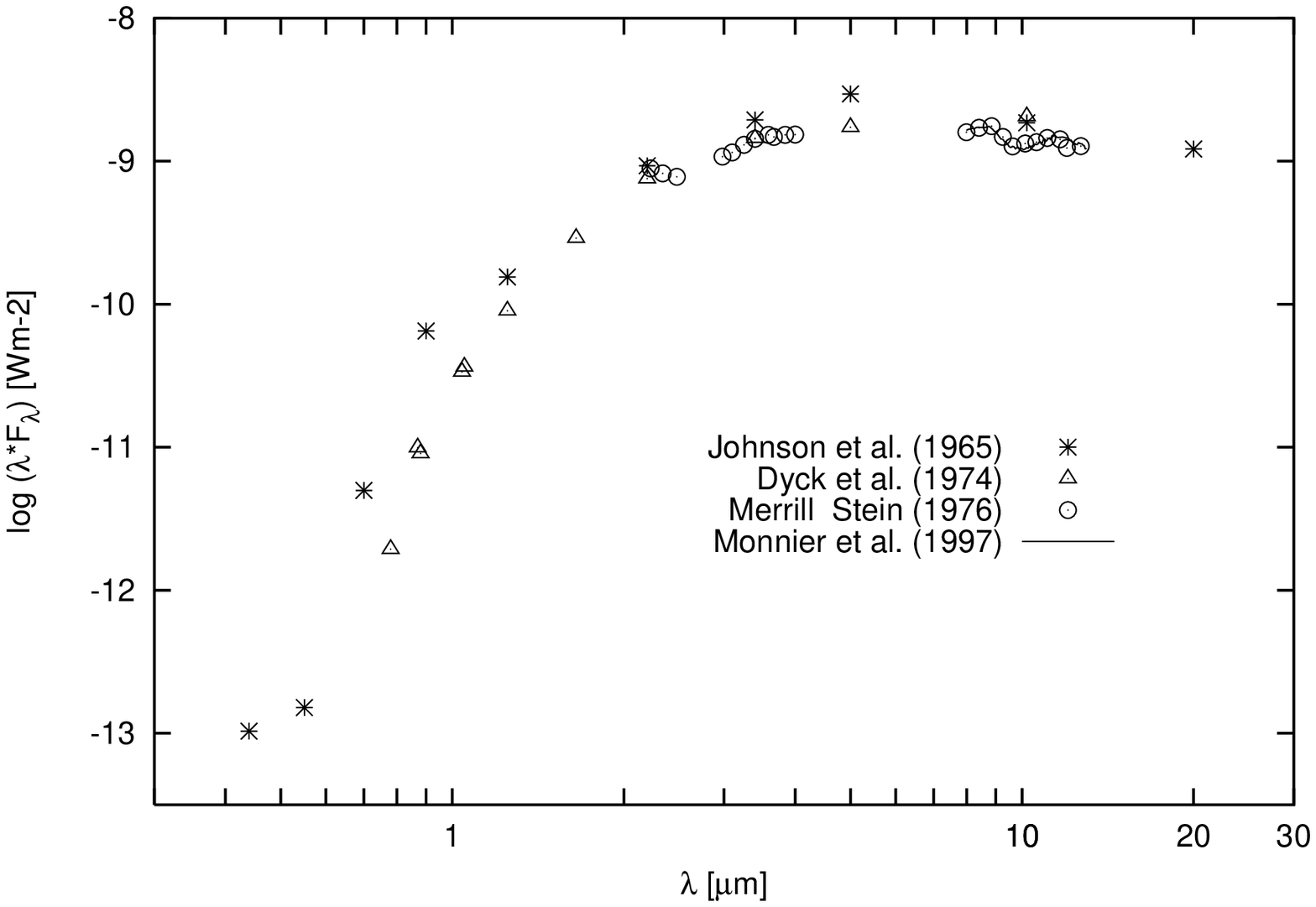}}
\epsfxsize=8.8cm
%%%% \mbox{\epsffile[40 50 547 410]{\RHOME/nmlcyg/aa2_sed_ground_iso.ps}}
\mbox{\epsffile[40 50 547 410]{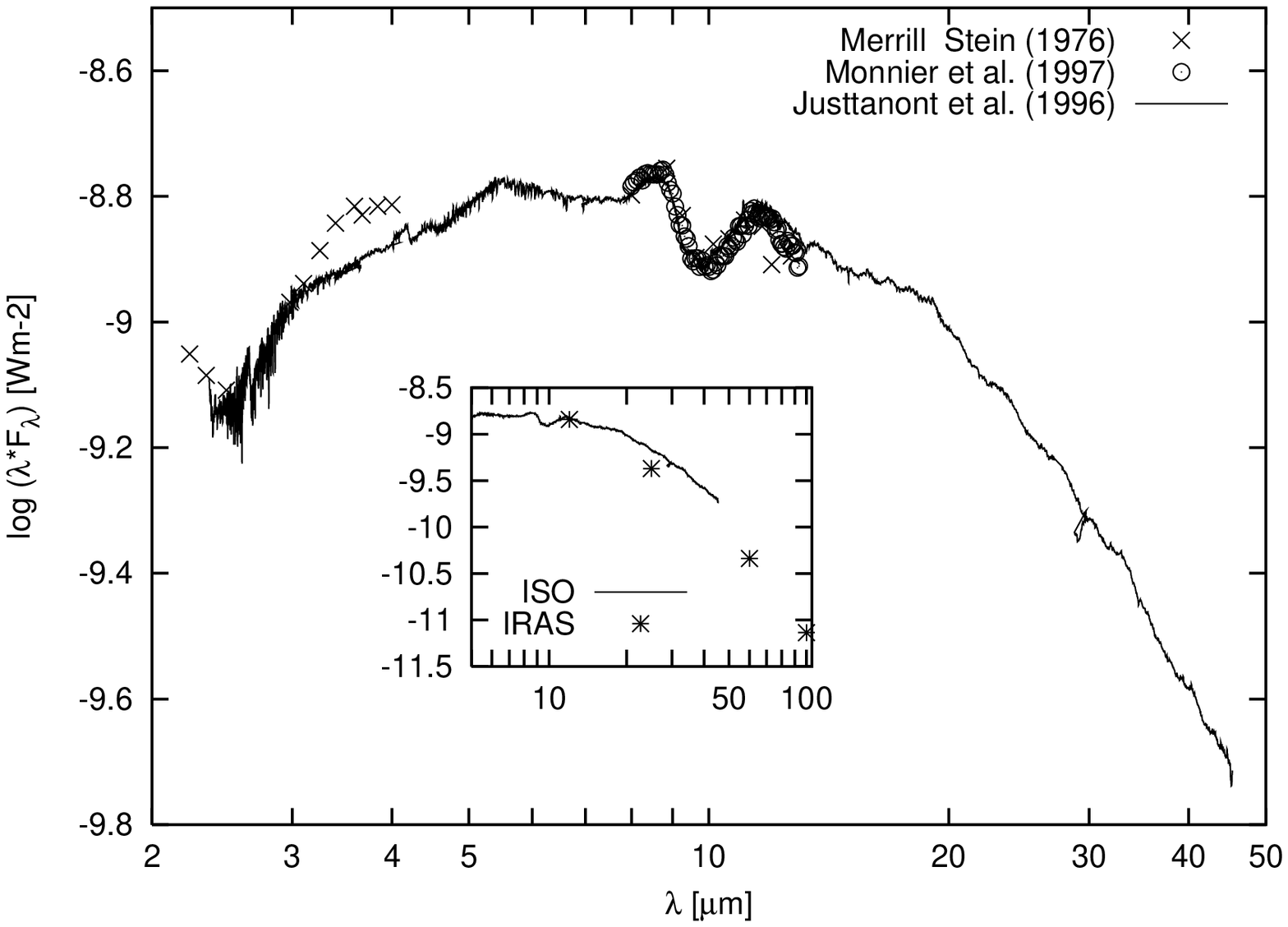}}
\caption[sed]
{Spectral energy distribution of \object{NML\,Cyg}
 (corrected for interstellar extinction of $A_{\rm v}$=3.7).
 {\bf Top:} Photometry of 
 Johnson et al. (1965), Dyck et al. (1974) and Merrill \& Stein (1976) as well
 as the UKIRT spectrophotometry of Monnier et al.\ (1997). The
 Merrill \& Stein (1976) data was scaled to match the UKIRT photometry.
 {\bf Bottom:} ISO-SWS spectrum of Justtanont et al.\ (1996),
 UKIRT spectrophotometry  of Monnier et al.\ (1997), and IRAS photometry.
 The ISO data was adjusted to match the ground-based photometry (see also
 Justtanont et al.\ 1996).
}                                      \label{Fsed}
%%%\protect{\vspace*{-0.1cm}}
\end{figure}
%%%%%%%%%%%%%%%%%%%%%%%%%%%%%%%%%%%%%%%%%%%%%%%%%%%%%%%%%%%%%%%%%%%%%%%%%%
%
Besides these ground-based observations \object{NML\,Cyg} has also been
observed by IRAS (Infrared Astronomical Satellite) in 1973  and
by ISO (Infrared Space Observatory) in 1996. %%%%%%%(\cite{JustaEtal96}).

\object{NML Cyg} is highly reddened
by the interstellar medium and the circumstellar shell. 
Low et al.\ (\cite{LowEtal70}) estimated
a total extinction
of $A^{\rm total}_{\rm V} \approx 9$\fm$ 5$ for \object{NML Cyg}. 
Based on radio continuum observations Gregory \& Seaquist (\cite{GreSea76}) 
obtained an interstellar extinction of $A_{\rm V} \approx 4$\fm$ 6$ whereas
Lee (\cite{Lee70}) estimated $E(B-V) = 1$\fm$ 2$
for the interstellar contribution.
We  will use the latter value of $E(B-V)=1$\fm$ 2$.
%%%%as in Monnier et al.\ (\cite{MonEtal97}).
This interstellar reddening was taken into account by 
adopting the method of Savage \& Mathis (\cite{SavMat79}) with
$A_{\rm V} = 3.1 E(B-V)$.

The spectral energy distrubion (SED) of \object{NML\,Cyg} is  shown in
Fig.~\ref{Fsed} and exhibits a 9.7\,$\mu$m silicate profile in absorption.
It includes the  (dereddened) data of Johnson et al. (1965), Dyck et al. (1974)
and Merrill \& Stein (1976) as well as the UKIRT spectrophotometry of
Monnier et al.\ (1997). Note that the  Merrill \& Stein (1976) data has been
scaled to match the UKIRT  spectrophotometry as in Monnier et al.\ (1997),
The spectral shapes of both data sets agree well. The lower panel of
Fig.~\ref{Fsed} shows the ISO-SWS spectroscopy (Justtanont et al.\ 1996)
in comparison with the UKIRT data and the IRAS photometry.
The ISO data was adjusted to match the ground-based photometry (see also
Justtanont et al.\ 1996).
In the long-wavelength region, the IRAS fluxes seem to be somewhat smaller
than those provided by ISO.
\section{Dust shell models}
%
%
%%%%%%%%%%%%%%%%%%%%%%%%%%%%%%%%%%%%%%%%%%%%%%%%%%%%%%%%%%%%%%%%%%%%%%%%%%%%
\begin{figure*}
\epsfxsize=12cm
%%%\mbox{\epsffile[52 193 555 728]{\RHOME/nmlcyg/aa3_sfv2_2500_td_1000_tau.ps}}
\mbox{\epsffile[52 193 555 728]{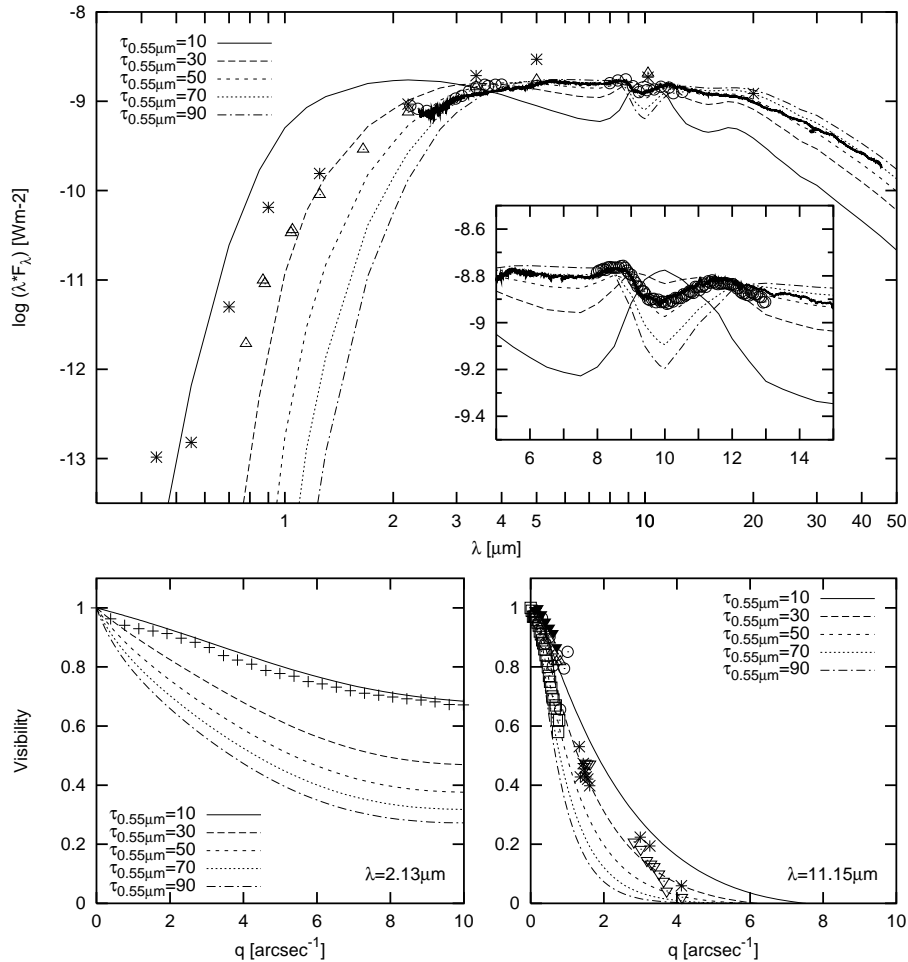}}
\hfill
\parbox[b]{5.5cm}{
\caption[sedtau]
{Optical depth dependence of
 model SED (top) and model
 visibilities at $2.13\,\mu$m (bottom left) and
 $11.15\,\mu$m (bottom right) for $T_{\rm eff}=2500$\,K
 and $T_{1}=1000$\,K. The  optical depth $\tau (0.55\mu m)$
 ranges between 10 and 90.
 The calculations are based on a black body, Ossenkopf et al.\ (1992)
 silicates, a Mathis et al.\ (1977) grain size distribution with
 $a_{\rm max}=0.25$, and a $1/r^{2}$ density distribution.
 Such a standard model does not appear to applicable to \object{NML\,Cyg}
 since different optical depths are required to match the various
 observations. 
 The symbols refer to the observations
 (see text, Figs.~\ref{Fvisi}, \ref{Fvisi11} and \ref{Fsed})
 corrected for interstellar extinction of $A_{\rm v}=3.7$:
\\
{\bf SED:}
 $\ast$: Johnson et al.\ (1965), $\bigtriangleup$: Dyck et al.\ (1974),
      $\circ$: Merril \& Stein (1976),
      thick line: Justtanont et al.\ (1996) [ISO data]. 
      The inlet shows the ISO data
      of Justtanont et al.\ (1996) (thick line) and the spectrophotometry
      of Monnier et al. (1997) (circles). \\
%{\bf \boldmath{$2.13\,\mu$}m visibility:} +: present paper. \\
%{\bf \boldmath{$11.15 \mu$}m visibility:}  
{\bf Visibilities:} +: $2.13\,\mu$m (present paper); \\
%{\bf \boldmath{$11.15 \mu$}m visibility:}  
  $\bigtriangleup$: 10.4\,$\mu$m (Dyck\,\&\,Benson 1992), 
  \ding{116}: 11.4\,$\mu$m (Dyck\,\&\,Benson 1994), 
  $\circ$: 11.2\,$\mu$m (Fix\,\&\, Cobb 1988), 
  $\ast$: 11.15\,$\mu$m (Monnier et al.\,(1997),  
  $\bigtriangledown$:  11.15\,$\mu$m Danchi et al.\ (1999),
  \ding{111}: 11.5\,$\mu$m Sudol et al.\ (1999).
\vspace*{5ex}
}                                      \label{Fsedtau}
}
\end{figure*}
%%%%%%%%%%%%%%%%%%%%%%%%%%%%%%%%%
%
In order to model both the observed SED and the $2.13\,\mu$m and 11.15\,$\mu$m
visibilities, radiative transfer calculations for the dust shell
were conducted. For the robust and non-ambiguous construction of dust shell
models it is essential to take diverse and independent observational
constraints into account. 
Apart from matching the spectral energy distribution, the 
consideration of spatially resolved information plays a crucial role for
obtaining a reliable model (see e.g.\ Bl\"ocker 1999).

\subsection{The radiative transfer code}
We used the code DUSTY developed by  Ivezi\'c et al.\ (\cite{IveNenEli97}), 
which solves the radiative transfer problem in spherical symmetry.
It considers absorption, emission and scattering and utilizes the
self-similarity and scaling behaviour of IR emission from radiatively
heated dust (Ivezi\'c \&  Elitzur \cite{IveEli97}).
To solve the radiative transfer problem
the spectral shape of the central source's radiation has to be specified and 
various properties of the surrounding dust shell  are required, viz. 
(i) the chemical composition and grain size distribution;
(ii) the dust temperature at the inner boundary;
(iii) the relative thickness, i.e. the ratio of outer to inner shell radius;
(iv) the  density distribution; and
(v) the total optical depth at a given reference wavelength.
The code has been expanded for the calculation of synthetic visibilities
as described by Gauger et al.\ (\cite{GauEtal99}).
\subsection{Single-shell models}
Dust shell models for \object{NML\,Cyg} were reported in
Rowan-Robinson \& Harris (1983), Rigdway et al.\ (1986), and
Monnier et al.\ (1997).  The first two studies used uniform outflow
models, i.e. approximated the density distribution of the dust shell by
$\rho \sim r^{-2}$. Monnier et al.\ (1997) found that uniform outflow
models appear not to be consistent with their near- and mid-infrared
observations and proposed a double-shell structure for the dust envelope.
It should be kept in mind, however, that the older  calculations are
partly based on different input quantities, as optical constants, grain
sizes, or shape of the central source radiation.

We calculated various uniform outflow models 
considering the following parameters within the radiative transfer
calculations: SED and visibilities were modelled for effective temperatures of
$T_{\rm eff}=2000$ to 3000\,K and
black bodies %%%%and Kurucz (1992) model atmospheres
as central sources of radiation.
We used the silicates of Ossenkopf et al.\ (\cite{OssEtal92}) which were found
to be well suited  to fit the silicate feature by Monnier et al.\ (1997). Some
test calculations were performed with the optical constants of 
of  Draine \& Lee (\cite{DraLee84}) and David \& Pegourie (\cite{DavPeg95})
as well.
For the grain-size distribution we considered
grain size distributions according to Mathis et al.\
(\cite{MRN77}, hereafter MRN),
i.e. $n(a) \sim a^{-3.5}$,
with  0.005\,$\mu {\rm m} \leq a  \leq (0.13$ to  $0.45)$\,$\mu$m   
as well as single-sized grains  with $a=0.05$ to 0.3\,$\mu$m.
%We used a   $1/r^{2}$
%density distribution
The shell thickness $Y_{\rm out} = r_{\rm out}/r_{1}$ amounted to 
$10^{3}$ to $10^{4}$
with $r_{\rm out}$ and $r_{1}$ being the outer and inner radius
of the shell, respectively. With these quantities specified,
the remaining fit parameters are the dust temperature at the inner boundary,
$T_{1}$, which, in turn, 
determines the radius of the shell's inner boundary, $r_{1}$,
and the optical depth, $\tau$, at a given reference wavelength,
$\lambda_{\rm ref}$. We refer to 
$\lambda_{\rm ref} =  0.55\,\mu$m. Models were calculated for
dust temperatures between 400 and 1500\,K and
optical depths between 10 and 100.

The neccessity to consider various observational constraints in order to
obtain dust-shell properties by radiative transfer calculations is
demonstrated in Fig.~\ref{Fsedtau}. 
It shows that the SED and the near-infrared and mid-infrared
visibilities cannot be fitted simultaneously by a given set of parameters.
Though each single observation can be matched reasonably well, different
values for the optical depth are required.
For example,  the K-band visibility requires  $\tau (0.55\mu m)=10$,
the N-band  visibility $\tau (0.55\mu m)=30$, and the silicate feature
$\tau (0.55\mu m)=50$ for the chosen dust-shell properties. High optical
depths ($\tau (0.55\mu m)\ga 30$) lead to a flux deficit in the optical.

This example refers to a given dust temperature of $T_{1} = 1000$\,K at the
inner boundary and an MRN grain-size distribution $n(a)$ with
$a_{\rm max} = 0.25\,\mu$m. Additionally the models rely on the assumption
of a standard density distribution $\rho(r) \sim 1/r^{2}$ whose validity
has been questioned by Monnier et al. (1997). 
Before abondaning the uniform outflow model, however,
one has to consider the effects of changes of
the former two dust-shell properties,  $T_{1}$ and  $n(a)$.

\begin{figure*}
\epsfxsize=12cm
%%%%\mbox{\epsffile[52 193 555 728]{\RHOME/nmlcyg/aa3_sfv2_2500_td_1000_tau30_gdmax.ps}}
\mbox{\epsffile[52 193 555 728]{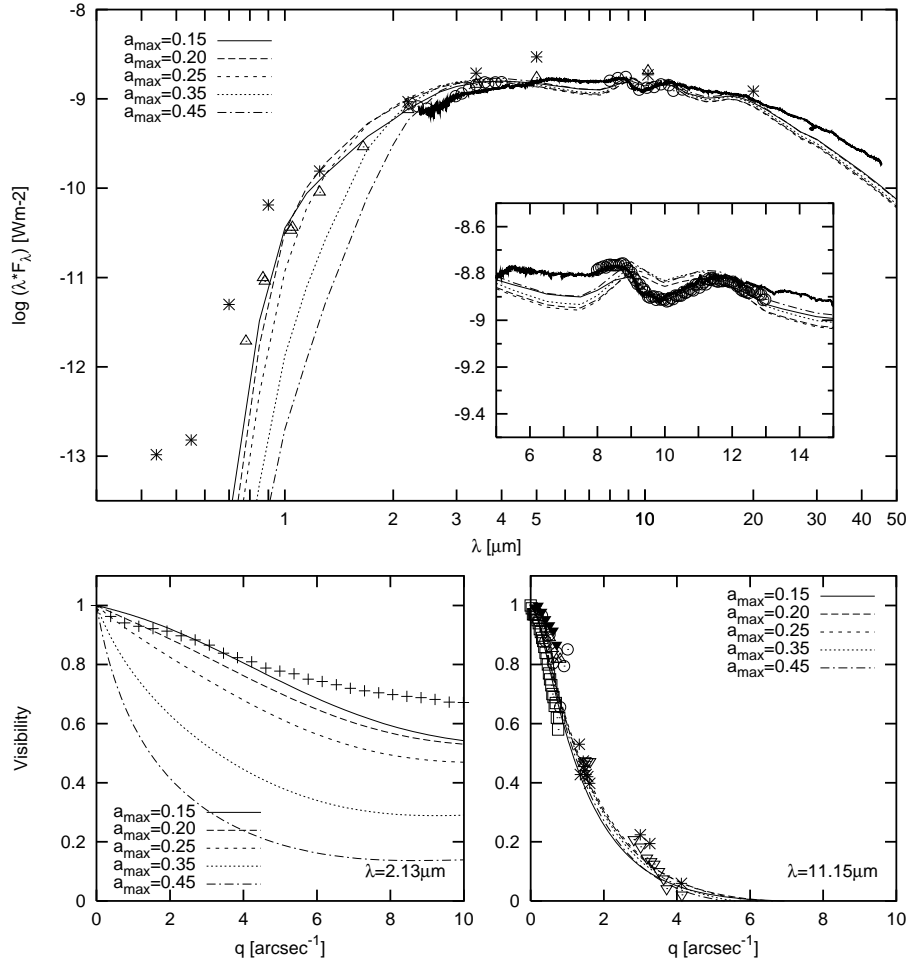}}
\hfill
\parbox[b]{5.5cm}{
\caption[gdmax]
{Maximum grain-size dependence
 of model SED (top) and model
 visibilities at $2.13\,\mu$m (bottom left) and
 $11.15\,\mu$m (bottom right) for $T_{\rm eff}=2500$\,K,
 $T_{1}=1000$\,K,  $\tau (0.55\mu m) = 30$.
 The maximum grain size $a_{\rm max}$ ranges between 0.15\,$\mu$m and
 0.45\,$\mu$m.
 The calculations are based on a black body, Ossenkopf et al.\ (1992)
 silicates, a Mathis et al.\ (1977) grain size distribution,
 $n(a) \propto a^{-3.5}$,
 and a $1/r^{2}$ density distribution.
 The symbols refer to the observations
 (cf.\ Figs.~\ref{Fvisi}, \ref{Fvisi11} and \ref{Fsed})
 corrected for interstellar extinction of $A_{\rm v}=3.7$
 and are the same as those in Fig.~\ref{Fsedtau}.
% \\
% {\bf SED:}
% $\ast$: Johnson et al.\ (1965), $\bigtriangleup$: Dyck et al.\ (1974),
%      $\circ$: Merril \& Stein (1976),
%      thick line: Justtanont et al.\ (1996) [ISO data]. 
%      The inlet shows the ISO data
%      of Justtanont et al.\ (1996) (thick line) and the spectrophotometry
%      of Monnier et al. (1997) (circles). \\
%{\bf Visibilities:} +: $2.13\,\mu$m (present paper); \\
%  $\bigtriangleup$: 10.4\,$\mu$m (Dyck \& Benson 1992),
%  $\blacktriangledown$:  11.4\,$\mu$m (Dyck \& Benson 1994),
%  $\circ$: 11.2\,$\mu$m (Fix \& Cobb 1988),
%  $\ast$: 11.15\,$\mu$m (Monnier et al.\ (1997),
%  $\bigtriangledown$  11.15\,$\mu$m Danchi et al.\ (1999).
%  $\Box$: 11.5\,$\mu$m Sudol et al.\ (1999).
}                                      \label{Fgdmax}
}
\end{figure*}

The $K$-visibility is very sensitive against scattering
and, thus, depends strongly on the assumed grain sizes.
Increasing the (maximum) grain size has a similar effect as increasing the
optical depth and results into a stronger decline of the K-visibility 
(cf.\ e.g.\ Bl\"ocker et al.\ 1999). This holds both for single-sized grains
and grain-size distributions. If one increases the maximum grain-size of
an MRN-like distribution most particles are still small due to the steep
decline of the distribution function towards larger grain sizes. However,
due to the very strong dependence of the absorption and scattering properties
on the grain sizes, these ``few'' larger particles determine
the  $K$-visibility equally as much as the bulk of the smaller grains
(Kr\"uger \& Sedlmayr 1997, Winters et al.\ 1997). On the other hand, at longer
wavelengths the changes induced by a larger maximum
grain size are predominantly of minor nature.
For instance, the SED silicate feature
and the mid-infrared visibility are only mildly affected. This fact allows,
in principle, to fix the optical depth by matching the silicate feature
and mid-infrared visibilities (if possible) and to fix independently the
grain-size distribution by the near-infrared data. The impact of
maximum grain sizes of MRN-like distribution on the various observational
quantities is illustrated in Fig.~\ref{Fgdmax}. These calculations show
that the low-frequency part of the $K$-band visibility may be fitted for
smaller optical depths if $a_{\rm max}$ is sufficiently small. However,
the silicate feature and the SED's long-wavelength tail are then only poorly
matched. The most favorable maximum grain size within an MRN-like distribution
was found to be close to $a_{\rm max} = 0.15\,\mu$m. 

To complete the discussion on grain sizes it has to be considered how
an MRN-like distribution compares with single-sized grains and which
choice is better suited for the instance of \object{NML\,Cyg}.
Whereas the $K$-visibility behaves similarly in both cases,
i.e. strongly declines with for increasing grain sizes, SED and mid-infrared
visibilities are more affected by variations of single-sized grains than by
those of a distribution. The interplay  between optical depth and grain size
is now much stronger. In summary, we found that single-sized grains appear
to give worse overall fits compared to grain-size distributions. 
%%%Though both cases do not lead to consistent dust-shell models
%%%,i.e.\  reasonable overall fits, with the current set of parameters, we 

%%%%%%%%%%%%%%%%%%%%%%%%%%%%%%%%%%%%%%%%%%%%%%%%%%%%%%%%%%%%%%%%%%%%%%%%%%%%
%%%%%%%%%%%%%%%%%%%%%%%%%%%%%%%%%%%%%%%%%%%%%%%%%%%%%%%%%%%%%%%%%%%%%%%%%%%%
%
%

Finally, we want to address the question of the dust temperature at the
inner shell boundary.
%%Fig.~\ref{Ftdust} shows the SED and visibilities for
%%$\tau(0.55\,\mu m)=30$ and $a_{\rm max} = 0.25 \mu$m.
The hotter the dust at the inner boundary of the shell,
the more flux is radiated in the
near-infrared and the less dominant is the long-wavelength tail of the
SED. Concomitantly, the visibilities decline at low spatial
frequencies the steeper, the hotter the inner dust-shell rim.  
Inspection of the model grid shows that the dust
temperature at the inner boundary should be close to 1000\,K.
Somewhat higher values, e.g.\ 1200\,K, are possible as well.

All parameter variations discussed so far
(of optical depths, dust temperatures, grain sizes etc.)
failed to give a consistent overall fit, thus the only basic parameter
left to vary is the density distribution (assuming spherical symmetry).
Consequently, it can be concluded
that the assumption of a uniform outflow model does not hold for
\object{NML\,Cyg} and that, for
instance, steeper density distributions or multiple shells have to be
considered. %% (provided the assumption of spherical symmetry is sufficient).
A similar conclusion was drawn by Monnier et al.\ (1997) who introduced
two shells with exponentially decreasing density distributions in order
to match the observations. 
\subsection{Multiple dust-shell components}
The evolution of massive supergiants can be accompanied by stages of episodic
mass-loss events as superwind phases or even mass-loss outbursts, as
for example, during the Red Supergiant Branch or the Luminous Blue Variable
stage (see, e..g., Vanbeveren et al.\ 1994, Heger et al.\ 1997,
de Jager \& Nieuwenhuijzen 1997, Langer et al.\ 1998).
During transitory phases
of heavy mass-loss the density in the dust shell is increased. This leads 
to regions in the dust shell with density enhancements over the normal
$1/r^{2}$ distribution. The radial density distribution may also change
within such superwind shells. For more details, see Suh \& Jones
(\cite{SuhJon97}). For example, such superwind models have 
successfully been applied to the hypergiant \object{IRC\,+10\,420} 
that is believed to have already moved off the RSG currently evolving into the
Wolf-Rayet stage (see Bl\"ocker et al.\ 1999).

\begin{figure}
\centering
\epsfxsize=7.0cm
%%% \mbox{\epsffile[95 199 566 755]{\RHOME/nmlcyg/aa3_swmodels_mult.ps}}
\mbox{\epsffile[95 199 566 755]{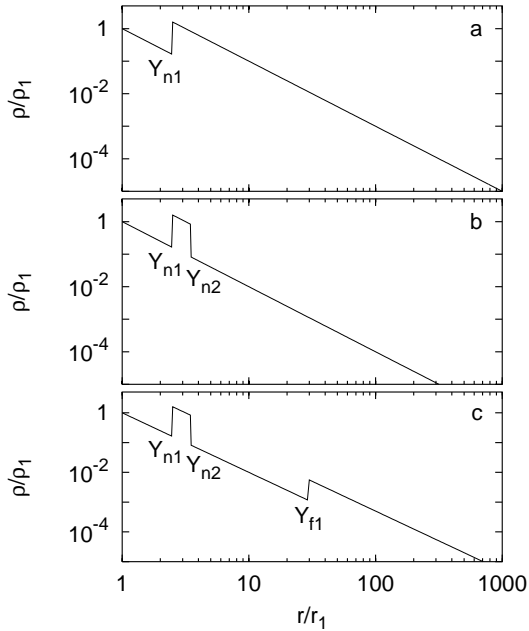}}
\caption[swmodel]{
Relative density distribution for different superwind models with
$\rho \sim 1/r^{2}$. The upper panel (a) shows a density enhancement
at $Y_{n1}=r_{n1}/r_{1}=2.5$, i.e. {\it n}earby the inner dust-shell rim,
with an amplitude of $A=10$.  The middle panel (b)
refers to a superwind shell ranging from $Y_{n1}=2.5$ to $Y_{n2}=3.5$
with $A=10$,
and the bottom panel (c) shows this  superwind shell and a additional
{\it f}ar-out density enhancement at  $Y_{f1}=30$ with $A=5$.
} \label{Fswmodel}
\end{figure}

Concerning the introduction of superwind models different degrees of complexity
can be considered: (i) single jumps with enhancement factors, or
amplitudes, $A$ at radii $Y_{n1}=r_{n1}/r_{1}$;
(ii) a single superwind shell with amplitude $A$ ranging from
$Y_{\rm n1}$ to $Y_{\rm n2}$; (iii) various shells; and
(iv) jumps or shells as in (i)-(iii) but with density distributions different
from the $1/r^{2}$ uniform outflow model. Examples of the superwind models
(i)-(iii) are illustrated in Fig.~\ref{Fswmodel}.
%%%%%%%%%%%%%%%%%%%%%%%%%%%%%%%
%%%%%%%%%%%%%%%%%%%%%%%%%%%%%%

In the following sections such superwind models will be discussed in more
detail.
In each case the suitability of other dust-shell parameters was carefully
checked. We found a optical depth of $\tau(0.55\mu {\rm m})$=30 to fit the
silicate feature. Within an MRN-like grain-size distribution a maximum
grain-size of $a_{\rm max}=0.15\mu$m was chosen in order to match the
overall pattern of the SED and the low-frequency curvature of the $K$-band
visibility. We stayed with the optical constants of Ossenkopf et al.\ (1992)
due to their excellent match of the silicate feature's shape. 
The dust-temperature at the inner dust-shell boundary amounted
to 1000\,K, the effective temperature of the central star to 2500\,K. 
\subsubsection{Two-component shells} \label{SSmultdust}
The most simple model of a previous epoch of enhanced mass-loss is that of  
a single density enhancement as shown in the upper panel
of Fig.~\ref{Fswmodel}. Calculations were conducted for relative
distances $Y_{n1}$ between 1.5 and 6.5 and amplitudes $A$ ranging from 
10 to 20. The best model found is that with $Y_{n1}=2.5$ and $A=10$ which
fits the SED from the optical to the infrared and matches the $K$-band
visibility reasonably well. The inner boundary of the dust shell amounts
to $r_{1}=5.8 R_{\ast}$ ($T_{1}=1000$\,K) corresponding to an
angular diameter of 95\,mas.
At $Y_{n1}=2.5$ the dust temperature has dropped to 621\,K.
More distant density jumps lead to worse fits of
the silicate feature and the far-infrared regime, less distant ones are not in
agreement with the $K$-band visibility (for given amplitude).
However, the mid-infrared visibility
was only poorly matched by all of these models giving too extended shells
at this wavelength. This, in turn,  indicates  that the cold dust-shell 
regions which determine the mid-infrared visibility have too high densities
within this  model.

\subsubsection{Embedded superwind shells}
\begin{figure*}
\epsfxsize=12cm
%%% \mbox{\epsffile[52 193 555 728]{\RHOME/nmlcyg/aa3_sfv2_2500_td_1000_tau_30_gdmax015_sw_2_5_10_ysh2_1.ps}}
\mbox{\epsffile[52 193 555 728]{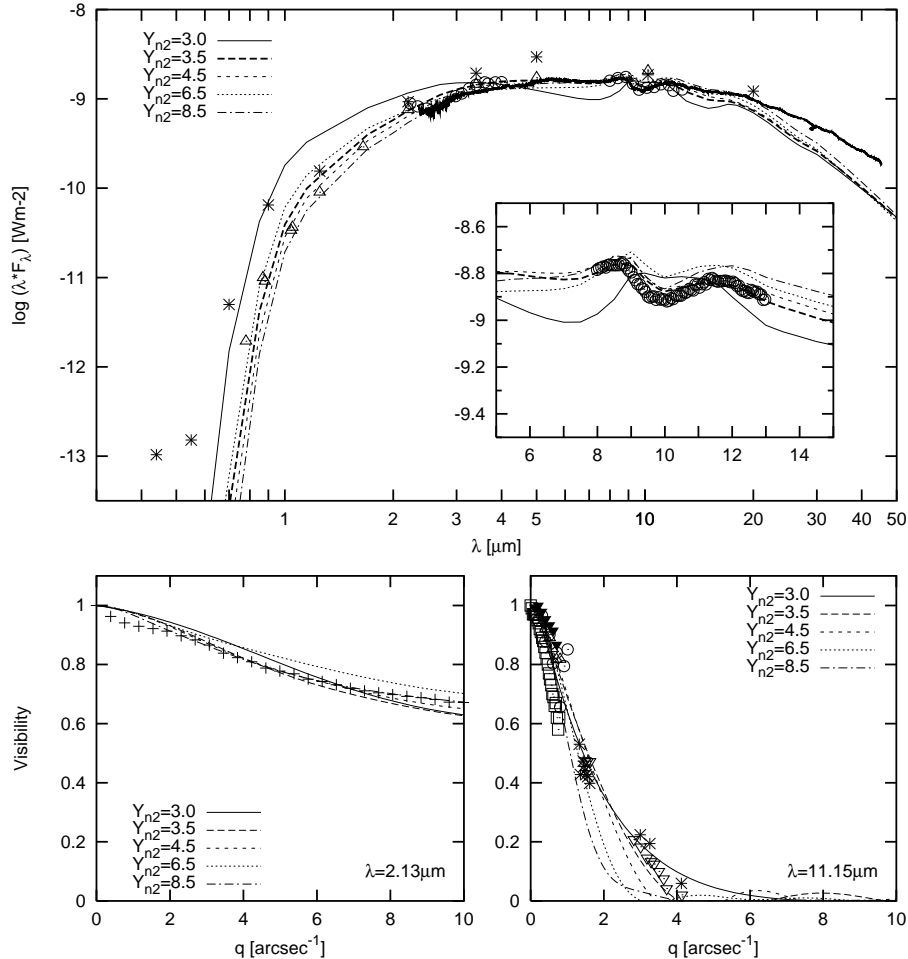}}
\hfill
\parbox[b]{5.5cm}{
\epsfxsize=5.5cm
%%%\mbox{\epsffile[96 83 566 247]{\RHOME/nmlcyg/aa3_swmodels_8b.ps}}
\mbox{\epsffile[96 83 566 247]{h2550f9.Sps}}
\caption[ysh2]
{Model SED (top, left) and model
 visibilities at $2.13\,\mu$m (bottom left)
 and $11.15\,\mu$m (bottom right) for $T_{\rm eff}=2500$\,K,
 $\tau (0.55\mu m)$=30, $T_{1}$=1000\,K, and different embedded
 superwind shells with 
 amplitude $A$=10 extending
 from $Y_{\rm n1}=2.5$ to  $Y_{\rm n2}=3.0$--8.5. The 
 density distribution is exemplified
 for $Y_{\rm n1}$=2.5 and $Y_{\rm n2}$=3.5
 in  the above panel (top, right)
 (cf.\ Fig.~\ref{Fswmodel}).
 The calculations are based on a black body, Ossenkopf et al.\ (1992)
 silicates, and a Mathis et al.\ (1977) grain size distribution
 with $a_{\rm max}$=0.15$\mu$\,m.
 The symbols refer to the observations
 (cf.\ Fig.~\ref{Fvisi}, \ref{Fvisi11} and \ref{Fsed})
 corrected for interstellar extinction of $A_{\rm v}$=3.7
 and are the same as those in Fig.~\ref{Fsedtau}.
%{\bf SED:}
% $\ast$: Johnson et al.\ (1965), $\bigtriangleup$: Dyck et al.\ (1974),
%      $\circ$: Merril \& Stein (1976),
%      thick line: Justtanont et al.\ (1996) [ISO data]. 
%      The inlet shows the ISO data
%      of Justtanont et al.\ (1996) (thick line) and the spectrophotometry
%      of Monnier et al. (1997) (circles). \\ 
%{\bf Visibilities:} +: $2.13\,\mu$m (present paper); \\
%  $\bigtriangleup$: 10.4\,$\mu$m (Dyck \& Benson 1992),
%  $\blacktriangle$:  11.4\,$\mu$m (Dyck \& Benson 1994),
%  $\circ$: 11.2\,$\mu$m (Fix \& Cobb 1988),
%  $\ast$: 11.15\,$\mu$m (Monnier et al.\ (1997),
%  $\bigtriangledown$  11.15\,$\mu$m Danchi et al.\ (1999).
%  $\Box$: 11.5\,$\mu$m Sudol et al.\ (1999).
}                                      \label{Fysh2}
}
\end{figure*}

Since the $K$-band visibility
traces the hot dust, it is sensitive to the place and strength of the density
jump but reacts only moderately to the extension of the density enhancement.
Consequently, the above model can be improved by introducing an embedded
high-density or superwind shell of given extension (see Fig.~\ref{Fswmodel},
middle panel) instead of a single jump. Since the inner boundary and the
amplitude of such a shell is strongly constrained by the $K$-band visibility, 
the inner boundary was fixed to $Y_{\rm n1}=2.5$ and the amplitude to $A=10$
within the corresponding model grid. Additional test calculations with other
values for  $Y_{\rm n1}$ and $A$ were performed as well. The outer boundary
$Y_{\rm n2}$ was varied between 3.0 and 10.5.
SED and visibility models for different values of $Y_{\rm n2}$ are shown in
Fig.~\ref{Fysh2}. It turned out that only a
close and comparatively thin shell ranging from  $Y_{\rm n1}=2.5$ to
$Y_{\rm n1}=3.5$ appears to be in agreement with both the visibility data at
different wavelengths and the SED from the optical to the mid-infrared.
Now, the inner boundary of the dust shell is located somewhat more outwards,
viz. at $R_{1}=6.2 R_{\ast}$ corresponding to an angular diameter of 101\,mas.
The superwind shell diameter extends from 252 to 353\,mas, the corresponding
dust
temperature at the inner and outer boundary amount to 587\,K to 411\,K, resp.
However, as obvious from
Fig.~\ref{Fysh2}
the model appears to be still incomplete due to a
lack of sufficiently high fluxes at longer wavelengths ($\lambda \ga 20\mu$m).
%%%%%%%%%%%%%%%%%%%%%%%%%%%%%%%%%%%%%%%%%%%%%%%%%%%%%%%%%%%%%%%%%%%%%%%%%%%%
%%%%%%%%%%%%%%%%%%%%%%%%%%%%%%%%%%%%%%%%%%%%%%%%%%%%%%%%%%%%%%%%%%%%%%%%%%%%
%
\subsubsection{Two superwind shells}
The above model of a thin density-enhanced region close to the inner
boundary of the dust shell gives a reasonable fit to the observed properties
from the optical to the mid-infrared. However, due to the small extension
of the embedded superwind shell, the densities in the outer regions of the
dust shell appear to be  too small to provide sufficiently high far-infrared
fluxes. This may indicate either the existence of a second, far-out density
enhancement or a shallower radial density dependence
(e.g. $\rho \sim 1/r^{1.5}$). In the case of a second episodic mass-loss
phase, the corresponding density enhancement
should be at moderate distances, close enough to raise the fluxes at
$\lambda > 20\mu$m but distant enough in order to preserve the
properties of the radiative transfer models at shorter wavelengths.

%%%%%%%%%%%%%%%%%%%%%%%%%%%%%%%%%%%%%%%%%%%%%%%%%%%%%%%%%%%%%%%%%%%%%%%%%%%%
\begin{figure*}
\epsfxsize=12cm
%%% \mbox{\epsffile[52 193 555 728]{\RHOME/nmlcyg/aa3_sfv2_2500_td_1000_tau_gdmax015_sw_2_5_10_3_5_1-30-5.ps}}
\mbox{\epsffile[52 193 555 728]{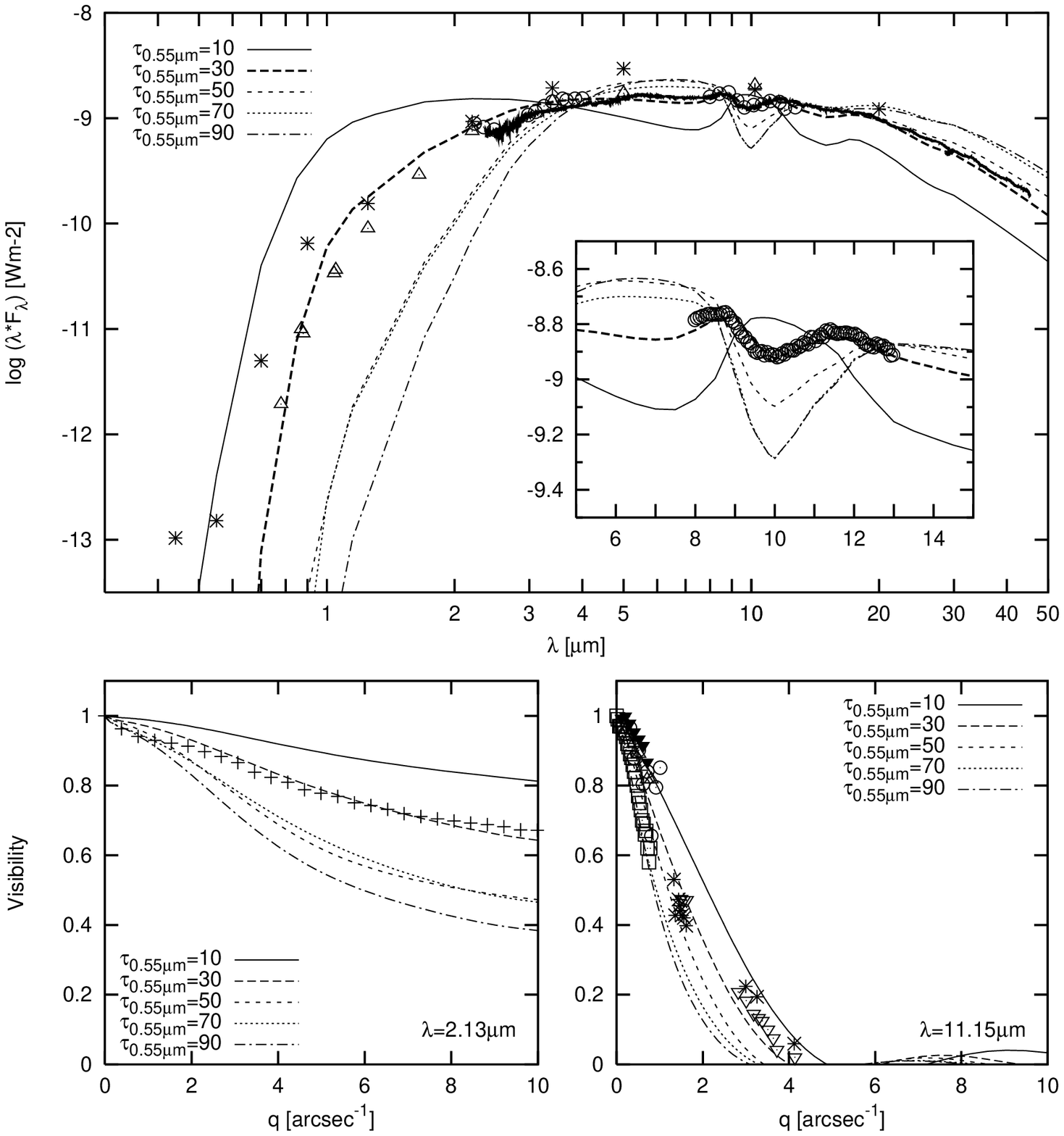}}
\hfill
\parbox[b]{5.5cm}{
\epsfxsize=5.5cm
%%%%\mbox{\epsffile[96 83 566 247]{\RHOME/nmlcyg/aa3_swmodels_8c.ps}}
\mbox{\epsffile[96 83 566 247]{h2550f10.Sps}}
\protect{\vspace*{-4mm}}
\caption[swmodelfar]
{Model SED (top) and model
 visibilities at $2.13\,\mu$m (bottom left) and
 $11.15\,\mu$m (bottom right) for an inner  superwind shell of
 amplitude $A$=10 between $Y_{\rm n1}$=2.5 and $Y_{\rm n2}$=3.5, and an
 additional far-out density enhancement at $Y_{\rm f1}$=30 with
 $A$=5. The density distribution is displayed in the above panel (top, right)
 (cf.\ Fig.~\ref{Fswmodel}).
 The models refer to
 $T_{\rm eff}$=2500\,K, $T_{\rm 1}$=1000\,K and 
 optical depths $\tau (0.55\mu m)$ ranging between 10 and 90.
 The calculations are based on a black body, Ossenkopf et al.\ (1992)
 silicates, and a Mathis et al.\ (1977) grain size distribution with
 $a_{\rm max}$=0.15$\mu$m.
 The symbols refer to the observations
 (cf.\ Fig.~\ref{Fvisi}, \ref{Fvisi11} and \ref{Fsed})
 corrected for interstellar extinction of $A_{\rm v}$=3.7
 and are the same as those in Fig.~\ref{Fsedtau}.
%{\bf SED:}
%  $\ast$: Johnson et al.\ (1965), $\bigtriangleup$: Dyck et al.\ (1974),
%      $\circ$: Merril \& Stein (1976),
%      thick line: Justtanont et al.\ (1996) [ISO data]. 
%      The inlet shows the ISO data
%      of Justtanont et al.\ (1996) (thick line) and the spectrophotometry
%      of Monnier et al. (1997) (circles). \\
%{\bf Visibilities:} +: $2.13\,\mu$m (present paper); \\
%  $\bigtriangleup$: 10.4\,$\mu$m (Dyck \& Benson 1992),
%  $\blacktriangle$: 11.4\,$\mu$m (Dyck \& Benson 1994),
%  $\circ$: 11.2\,$\mu$m (Fix \& Cobb 1988),
%  $\ast$: 11.15\,$\mu$m (Monnier et al.\ (1997),
%  $\bigtriangledown$  11.15\,$\mu$m Danchi et al.\ (1999).
%  $\Box$: 11.5\,$\mu$m Sudol et al.\ (1999).
}                                      \label{Fswmodelfar}
}
\end{figure*}
%%%%%%%%%%%%%%%%%%%%%%%%%%%%%%%%%%%%%%%%%%%%%%%%%%%%%%%%%%%%%%%%%%%%%%%%%%%%

Grids were calculated for such an additional far-out superwind shells
with an inner boundary located between $Y_{\rm f1}=7$ to 50 and amplitudes
$A$ ranging between 2 and 50.
The best model found was that with
$Y_{\rm f1}=30$ and a moderate amplitude of $A=5$.
For simplicity we considered here only
density jumps as shown in Fig.~\ref{Fswmodel}. Note that
though the existence of such a {\rm distant} density enhancement appears to be
well constrained by the observations within the current model, the
corresponding properties of the distant shell are less constrained.
For instance, different combinations of location,
amplitude and density slope can lead to models of comparable quality.
Hovever, shells at significantly smaller or larger distances can clearly be
excluded.

Since the introduction of a far-out density enhancement hardly affects the
inner regions, the inner boundary of the dust shell keeps at 6.2 stellar
radii corresponding to an  angular diameter of
105\,mas. The angular diameter increased somewhat compared to the case without
an outer shell due to the the (slight) increase of the bolometric flux.
The diameter of the embedded close superwind shell extends from
263\,mas to 368\,mas, and the inner boundary of the distant shell has a
diameter of 3\farcs 15. The dust temperature within
the embedded superwind shell ranges from 584\,K to 430\,K and within the
distant shell from 149\,K to 40\,K. 

\begin{figure}
\centering
\epsfxsize=8.8cm
\centering
%%% \mbox{\epsffile{\RHOME/nmlcyg/aa3_ffraction_2500_td_1000_tau_30_gdmax015_sw_2_5_10_3_5_1-30-5.ps}}
\mbox{\epsffile{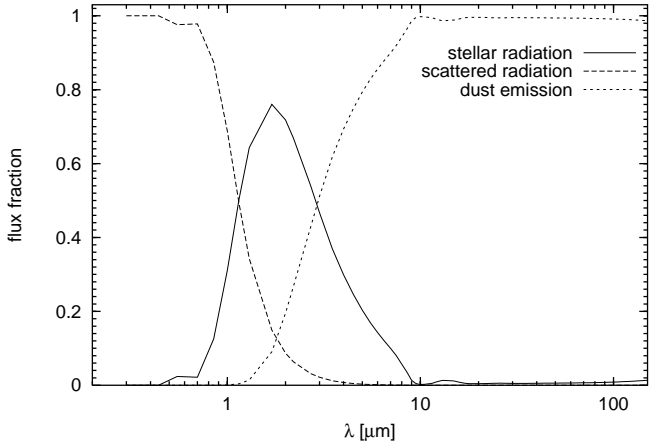}}
\caption[ffraction]
{Fractional contributions of the emerging stellar radiation 
 as well as  of the scattered radiation and of the dust emission to the total
 flux as a function of the wavelength for the model with
 for an inner  superwind shell of
 amplitude $A$=10 between $Y_{\rm n1}$=2.5 and $Y_{\rm n2}$=3.5 and an
 additional far-out density enhancement at $Y_{\rm f1}$=30 with
 $A$=5 (see Fig.~\ref{Fswmodelfar}).
 Model parameters are:
 black body, $T_{\rm eff}=2500$\,K, $T_{1}=1000$\,K,
 $\tau_{0.55\mu{\rm m}}=30$, Ossenkopf et al.\ (1992) silicates, and a
 Mathis et al.\ (1977) grain size distribution with $a_{\rm max}=0.15\,\mu$m.
}                                      \label{Fffraction}
\end{figure}

Fig.~\ref{Fffraction} gives the fractional flux contributions of the
emerging stellar radiation, of the scattered radiation and of the
thermal dust emission as a function of wavelength for this model.
At $2.2\mu$m, the flux is mainly determined by direct stellar light (67\%).
Scattered radiation and dust emission contribute 6\% and 27\%, resp.
Thus, roughly 80\%  of the circumstellar shell's $K$-band radiation is due to
thermal dust emission. At wavelengths larger than 10$\mu$m
dust emission completely dominates. The bolometric flux  amounts to
$F_{\rm bol} = 3.63 \cdot 10^{-9}$\,Wm$^{-2}$.
Accordingly, the central star
%%%%has a luminosity of $L/L_{\odot} = 113\,129 \cdot (d/{\rm kpc})^{2}$ and
has a luminosity of
$L/L_{\odot} = 1.13 \cdot 10^{5} \cdot (d/{\rm kpc})^{2}$
and an angular diameter of
$\Theta_{\ast} = 1.74 \cdot 10^{9} \,
                 \sqrt{ F_{\rm bol}/T_{\rm eff}^{4} } \sim 16.2$\,mas.

\begin{figure}
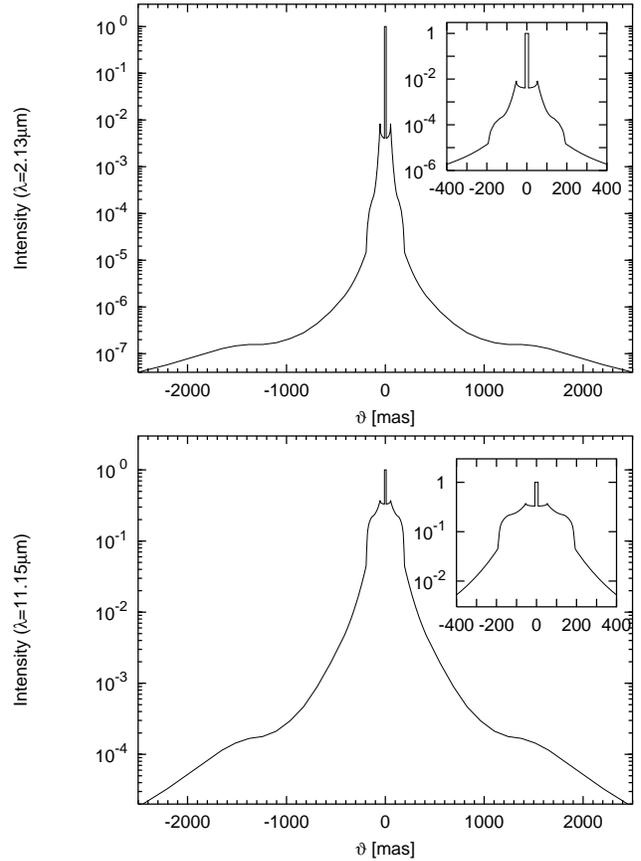

\centering
\epsfxsize=8.8cm
%%% \mbox{\epsffile[49 432 570 770]{\RHOME/nmlcyg/aa3_intensity_2mu_2500_td_1000_tau_30_gdmax015_sw_2_5_10_3_5_1-30-5.ps}}
\mbox{\epsffile[49 432 570 770]{h2550f12.Tps}}
\epsfxsize=8.8cm
%%% \mbox{\epsffile[49 432 570 770]{\RHOME/nmlcyg/aa3_intensity_11mu_2500_td_1000_tau_30_gdmax015_sw_2_5_10_3_5_1-30-5.ps}}
\mbox{\epsffile[49 432 570 770]{h2550f12.Bps}}
\caption[intensity]
{Normalized intensity at $2.13\,\mu$m (top) and   $11.15\,\mu$m (bottom)
 vs.\ angular displacement $\vartheta$
 for a superwind model with  an inner embedded superwind shell of
 amplitude $A=10$, located between $Y_{\rm n1}=2.5$ and $Y_{\rm n2}=3.5$,
 and an additional far-out density enhancement at $Y_{\rm f1}=30$ with
 $A=5$. The central peak belongs to the central star.
 The inner hot rim of the circumstellar shell has a radius of 52.5\,mas,
 the inner embedded superwind shell extends from 131.5\,mas to 184\,mas, and  
 the cool far-out component is located at a radius of 1500\,mas.
 All loci correspond to local intensity maxima or plateaus. In the $K$-band
 the inner rim of the circumstellar shell is noticeably limb-brightened.
 The models refer to
 $T_{\rm eff}=2500$\,K, $T_{1}=1000$\,K, and
  $\tau (0.55\mu m) = 30$.
 The calculations are based on a black body, Ossenkopf et al.\ (1992)
 silicates, and a Mathis et al.\ (1977) grain size distribution with
 $a_{\rm max}=0.15\mu$m.
}                                      \label{Fintensity}
\end{figure}

Fig.~\ref{Fintensity} shows the normalized intensity distribution at
$2.13\mu$m (upper panel) and $11.15\mu$m (lower panel) as
function of angular distance. The barely resolved central peak corresponds
to the central star
(angular diameter: 16.2\,mas). At $2.13\mu$m the inner boundary
of the dust shell (radius: 52.5\,mas) is limb-brightened resulting in a
ring-like intensity distribution. The optical depth
($\tau(2.13\mu{\rm m})=1.5$) is still moderate leading to 
a noticable effect of a limb-brightened dust condensation zone 
(cf.\ Ivezic \& Elitzur 1996,  Bl\"ocker et al.\ 1999).
In the $N$ band ($\tau(11.5\mu{\rm m})=2.2$) this is hardly visible. 
The embedded superwind shell leads to an increase
of the intensity distribution's waist at angular radii between 131.5\,mas and
184\,mas. Finally, the far-out shell provides broad low-intensity wings
at angular radii larger than $\sim 1\farcs 5$.

Asuming an outflow velocity of 25\,km/s and a distance of 1.8\,kpc,
the kinematical ages of these
shells can be calculated. The inner embedded shell corresponds to a phase
of enhanced mass-loss lasting for $\sim 18$\,yr
in the immediate history of \object{NML\,Cyg}. It began 59.2\,yr ago and
ceased 41.5\,yr ago. During this time span the mass-loss rate
increased by a factor of 10. Correspondingly, the outer shell, i.e. the
density jump at $Y_{f1}=30$, is due to
to a high mass-loss period which terminated 529.3\,yr ago. Throughout this
phase in the past matter was expelled with a fivefold increased rate. 
With a dust-to-gas ratio of 0.005 and a specific dust density of
3\,g\,cm$^{-3}$ the model gives a present-day mass-loss rate of
$\dot{M} = 1.2 \cdot 10^{-4}$\,M$_{\odot}$/yr which is in line with the
CO and OH observations (see Sect.~\ref{Sintro}).
%% d(stern)=16.4mas Mdust=6.0956e-07 Msol/yr
%%Kinematical Time Scales: \\
%%inner shell: 41.5 yr to 59.2   with Vexp=25 and d=1.8kpc \\
%%%outer shell: 529.3 yr 

%%%%\subsubsection{The density distribution in the outer shell}
\subsubsection{Other density slopes}
In the above sections the uniform-outflow model was abandoned due to its
mismatch of the observed properties. The introduction of two superwind shells,
or density enhancements of certain amplitude $A$,
leads to a model which fits the observations from the optical to the infrared.
However, so far it was assumed that $\rho \sim 1/r^{2}$ holds within
each dust-shell region.
Assuming the outflow velocity kept constant any shallower (steeper)
density slope than  $\rho \sim 1/r^{2}$ means that the mass-loss rate has
decreased (increased) with time in the respective dust-shell regions.
%
%The SED shown in
%{\bf Fig.~\ref{Fswmodelfar}}
%indicates that the double-shell
%superwind models still seem to predict a somewhat larger spectral index,
%i.e.\ faster flux decrease, than the ISO observations show for
%$\lambda \ga 20$.
%Additional models with shallower density distributions in the outer shell
%were calculated to improve the model. The grid comprises density distributions
%from $\rho \sim 1/r^{2}$ to  $\rho \sim 1/r^{1.5}$  for $Y \ge 30$.
%Fig.~\ref{Fswmodeldens}
%shows the results of these calculations in comparison with
%the observations in the case of the SED. Additionally, the $400 \mu$m
%and 1.3\,mm data of Sopka et al.\ (1985) and Walmsley et al.\ (1991), resp.,
%were included. A somewhat shallower distribution of
%$\rho \sim 1/r^{1.7}$  in the far-out regions of the dust shell appears to be
%slightly better suited than the standard model with $\rho \sim 1/r^{2}$.
%The other properties of the dust shell as, e.g., visibilities
%at 2.13$\mu$m and
%$11.15\mu$m are scarcely affected. Larger spectral indices,
%as e.g.\ 1.5, overestimate
%the infrared fluxes and impair the visibility fits as well. The 1.3\,mm
%data, however, cannot be matched by any of the models.
%It has already been noted by  Sopka et al.\ (1985) and Walmsley et al.\ (1991)
%that the fluxes at 400\,$\mu$m and 1.3\,mm are in excess of standard models.
%It has been argued that flat density distributions (as introduced above) or
%an increased emissivity due to amorphous silicates may be responsible for
%this excess. 

The double-shell
superwind models still seem to predict a somewhat larger spectral index,
i.e.\ faster flux decrease, than the ISO observations indicate for
$\lambda \ga 20$ (Fig.~\ref{Fswmodelfar}), and than
$400 \mu$m and 1.3\,mm data of Sopka et al.\ (1985) and
Walmsley et al.\ (1991), resp., show.
Additional models with shallower density distributions in the outer shell
were calculated to improve the model.
A somewhat shallower distribution of
$\rho \sim 1/r^{1.7}$  in the far-out regions ($Y \ge 30$)
of the dust shell appears to be
slightly better suited than the standard model with $\rho \sim 1/r^{2}$ and
matches also the  400\,$\mu$m observations. 
The 1.3\,mm data, however, cannot be matched by any of the models.
It has already been noted by  Sopka et al.\ (1985) and Walmsley et al.\ (1991)
that the fluxes at 400\,$\mu$m and 1.3\,mm are in excess of standard models.
It has been argued that flat density distributions (as introduced above) or
an increased emissivity of the silicates may be responsible for
this excess.

%%%\subsubsection{Other density slopes in general}
If one considers different density slopes in the outer shell to improve the
model, the question arises if this should not be considered for the whole
dust shell model in general. For instance, Monnier et al.\ (1999) found
two shells with exponentially decreasing density distributions appropriate
to match the observations. In order to prove (or disprove)
the applicability of the assumed $1/r^{2}$ distribution in the inner regions
of the shell ($Y\le 30$), we re-calculated the main body of the grid presented
so far including the superwind shells for density distribution ranging from
$\rho \sim 1/r$ to $\rho \sim 1/r^{7}$ and combinations of it.
We found the same best model as presented above to match the observations from
the optical to the sub-mm domain, i.e. a (close-to) $1/r^{2}$ overall
density distribution with a thin superwind shell
($Y=2.5$ to 3.5) near to the inner dust-shell rim
and a far-out density enhanced region ($Y\ga 30$).
Shallower distributions in the inner region can be excluded, for instance
due to the mismatch of the mid-infrared visibility. Steeper distributions
partly match the observations but were found to be unable to provide an 
overall fit. If one considers only the embedded superwind shell, models
can be constructed which show, in principle, similar properties as
the one presented above but have a much steeper density distribution within
the embedded superwind shell,
i.e. as steep as $1/r^{5-7}$. Then the superwind shell
has to be  somewhat more extended and to exhibit a larger amplitude
(e.g.\ $A=20$).
However, preference can be given to the  model 
showing superwind shells of enhanced density but preserving a $1/r^{2}$
distribution in the various parts of the dust shell due to a better overall
match of the observations. 
%%%%%%%%%%%%%%%%%%%%%%%%%%%%%%%%%%%%%%%%%%%%%%%%%%%%%%%%%%%%%%%%%%%%%%%%
%%%%%%%  Model fits: flux contributions  for two shells
%%%%%%%%%%%%%%%%%%%%%%%%%%%%%%%%%%%%%%%%%%%%%%%%%%%%%%%%%%%%%%%%%%%%%%%%%%%%%%
%%%%%%%%%%%%%%%%%%%%%%%%%%%%%%%%%%%%%%%%%%%%%%%%%%%%%%%%%%%%%%%%%%%%%%%%%%%%%%%
%%%%%%%%%%%%%%%%%%%%%%%%%%%%%%%%%%%%%%%%%%%%%%%%%%%%%%%%%%%%%%%%%%%%%%%%
%%%%%%%  Model fits: Intensity models 
%%%%%%%%%%%%%%%%%%%%%%%%%%%%%%%%%%%%%%%%%%%%%%%%%%%%%%%%%%%%%%%%%%%%%%%%%%%%%%
%%%%%%%%%%%%%%%%%%%%%%%%%%%%%%%%%%%%%%%%%%%%%%%%%%%%%%%%%%%%%%%%%%%%%%%%%%%%%%%
%
%
\section{Summary and discussion}
The first diffraction-limited
$2.13\,\mu$m observations of \object{NML\,Cyg} with
73\,mas resolution were presented.
The speckle interferograms were obtained with the 6\,m
telescope at the Special Astrophysical Observatory, and the
image reconstruction was based on the bispectrum speckle-interferometry method.
In order to interpret these observations 
radiative transfer calculations have been carried out.
We modelled the spectral energy distribution,
in particular the ISO LWS spectroscopy (Justtanont et al.\ 1996),
the $2.13\,\mu$m visibility function, and 
mid-infrared visibility functions at $11\,\mu$m
(e.g.\ Danchi et al.\ 1999). Preference was given to those observations
conducted in phase or close to the $2.13\,\mu$m observations in June 1998. 

The observed dust shell properties do not appear to be in accordance with
standard single-shell models (uniform outflow) but seem to require
multiple components. Since single-shell models fail to match the observations,
density enhancements in the dust shell were considered. Such density
enhancements correspond to previous periods of enhanced mass-loss.
The  episodic and/or stochastic occurence of such superwind phases seems to be
a typical feature of  the late stages of massive  star evolution. 
An extensive grid of models was calculated for different
locations and amplitudes of such superwind regions in the dust shell.
It turned out that at least two superwind phases have to be taken into account
in order to to match the observations from the optical to the
sub-mm domain. 
The best model is that of  a dust shell with a temperature of 1000\,K at its
inner radius of $6.2 R_{\ast}$. It includes a close embedded superwind shell
extending from  $15.5 R_{\ast}$ to  $21.7 R_{\ast}$ with an amplitude of 10
and a far-out density enhancement at  $186 R_{\ast}$ with an amplitude of 5.
The inner rim of the dust shell, i.e. the dust condensation zone, has a 
dust temperature of $T_{1}=1000$\,K and an angular diameter of  105\,mas.
At  $2.13\,\mu$m this zone is limb-brightened leading to a ring-like 
intensity distribution.
The diameter of the embedded close superwind region extends from
263\,mas to 368\,mas and is composed of relatively cool dust with
temperatures ranging from 583 to 430\,K.  
At the inner boundary of the distant superwind region the dust temperature
has dropped to 149\,K corresponding to an inner diameter of 3\farcs 15.

\begin{figure}
\centering
\epsfxsize=8.8cm
%%% \mbox{\epsffile{\RHOME/nmlcyg/aa3_visi_2.2_swlong.ps}}
\mbox{\epsffile{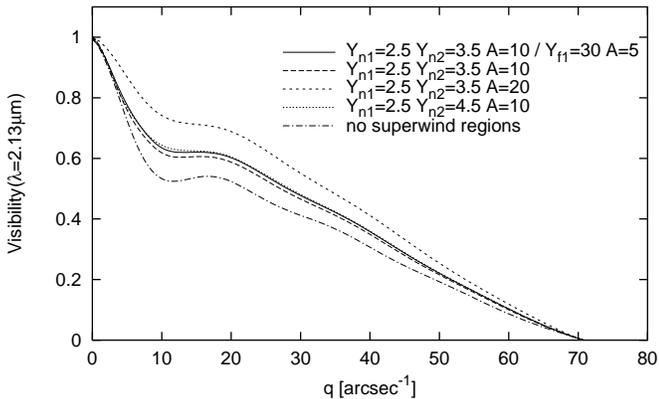}}
\caption[vis_2.2_swlong]
{Model visibilities at $2.13\mu$m in the long-baseline
 inter\-fero\-metry range up to baselines of 36m
 (6\,m:  $q$=13.4 arc\,sec$^{-1}$)
 for different assumptions on the density
 distribution. The solid line gives the best fit to the observations and
 refers to a dust shell with an  inner embedded superwind region with
 amplitude $A=10$, located between $Y_{\rm n1}=2.5$ and $Y_{\rm n2}=3.5$,
 and an additional far-out density enhancement at $Y_{\rm f1}=30$ with
 $A=5$. The other lines correspond to models without an outer density
 enhancement and a superwind shell of equal properties, of larger amplitude,
 and larger extension, resp. A standard model without any density enhancement
 is given as well.
% The visibilities are shown in the long-baseline
% interferometry range up to baselines of 36m
% (6\,m:  $q$=13.4 arc\,sec$^{-1}$).
 The models refer to
 $T_{\rm eff}=2500$\,K, $T_{1}=1000$\,K,  and $\tau (0.55\mu m)=30$
  and are based on a black body, Ossenkopf et al.\ (1992)
 silicates, and a Mathis et al.\ (1977) grain size distribution with
 $a_{\rm max}=0.15\mu$m.
% The symbols refer to the observations
% (see text, Fig.~\ref{Fvisi}, \ref{Fvisi11} and \ref{Fsed})
% corrected for interstellar extinction of $A_{\rm v}=3.7$. \\
}                                      \label{Fvisi_2.2_swlong}
\end{figure}

Fig.~\ref{Fvisi_2.2_swlong} shows the 2.13\,$\mu$m visibility for baselines
up to 36\,m and different assumptions on the superwind regions in
comparison with a standard uniform outflow model. The central star, having
a diameter of $\sim 16$\,mas, becomes partially
resolved for baselines larger than 10\.m and is completely resolved at a
baseline of 31\,m.  
%%%%%%%%%%%%%%%%%%%%%%%%%%%%%%%%%%%%%%%%%%%%%%%%%%%%%%%%%%%%%%%%%%%%%%%%%%%%
%%%%%%%%%%%%%%%%%%%%%%%%%%%%%%%%%%%%%%%%%%%%%%%%%%%%%%%%%%%%%%%%%%%%%%%%%%%%

During the entire grid calculations the validity of other parameters as,
optical depths, dust temperatures, grain sizes, effective temperatures, etc.,
was
comprehensively
checked. The grain sizes, $a$, were found to be in accordance with
a standard distribution function,
$n(a)$\,$\sim$\,$a^{-3.5}$, with $a$ ranging between
$a_{\rm min}$\,=\,0.005\,$\mu$m and $a_{\rm max}$\,=\,0.15\,$\mu$m.
The effective temperature was chosen to be 2500\,K. The temperature of a giant
with spectral type M6\,III is $\sim 3200$\,K (Perrin et al.\ 1998) whereas
supergiants (I-II) have systematically lower temperatures
(Dyck et al.\ 1996, 1998). For an M6 supergiant the effective temperature
can be estimated to be close to 3000\,K. However, (moderate) variations of the
effective temperature affect the radiative transfer models only slightly.
Thus, though a somewhat higher effective temperature
appears to better comply with the spectral type,
differences to the presented $T_{\rm eff}=2500$\,K
grid are only of minor nature
and results and conclusions of the presented study remain unchanged.
Corresponding test calculations have been performed for
$T_{\rm eff}=3000$\,K and $T_{\rm eff}=3500$\,K. The temperature at the inner
rim of the dust shell was found to be more or less fixed at 1000\,K
in order to be 
in accordance with the observations. However, slightly higher temperatures,
e.g.\ 1100\,K, appear to be possible as well.
The bolometric flux  amounts to $F_{\rm bol} = 3.63 \cdot 10^{-9}$\,Wm$^{-2}$
corresponding to a central-star 
%%%%luminosity of $L/L_{\odot} = 113\,129 \cdot (d/{\rm kpc})^{2}$.
luminosity of
$L/L_{\odot} = 1.13 \cdot 10^{5} \cdot (d/{\rm kpc})^{2}$.
Accordingly, \object{NML\,Cyg}  is a true massive supergiant with
$L \sim 4 \cdot 10^{5} L_{\odot}$ at the distance of the  Cyg OB2 association. 

Though the density distribution was scaled by amplitudes $A$ in various
parts of of the dust shell, its general $1/r^{2}$ dependence
proved to be appropriate to model the observations and thus could be
maintained. Alternative models
%%%%%of somewhat minor quality
were found if one considers a much steeper density distribution
in the inner embedded
superwind region. Then, however, the superwind shell has the same inner 
diameter but a larger amplitude and extension. 
A slight improvement of the far-infrared properties can be obtained if
a shallower density distribution of $\rho \sim 1/r^{1.7}$ is considered
in the distant superwind region. Then, assuming that the outflow velocity
kept constant, the mass-loss rate in the outer component has decreased with
time.
The present-day mass-loss rate can be determined to be
$\dot{M} = 1.2 \cdot 10^{-4}$\,M$_{\odot}$/yr.
The inner embedded superwind shell corresponds to a phase
of enhanced mass-loss ($\dot{M} = 1.2 \cdot 10^{-3}$\,M$_{\odot}$/yr)
in the immediate history of \object{NML\,Cyg}
which began 59.2\,yr ago and lasted for $\sim 18$\,yr
if $d=1.8$\,kpc and $v=25$\,km/s.
Correspondingly, the outer superwind region is due to
to a high mass-loss period ($\dot{M} = 6.0 \cdot 10^{-4}$\,M$_{\odot}$/yr)
which terminated 529.3\,yr ago.

The inner diameter of the dust shell of 105\,mas is close to the one  found
by the standard uniform-outflow models of Rowan-Robinson \& Harris (1983)
and Ridgway et al.\ (1986) (80\,mas and 90\,mas, resp.)
but only half as large as the one found by Monnier et al.\ (1997) (240\,mas).
Note, however, that the repsective models differ in various aspects.
On the one hand, the simultaneous match of SED and visibilities appear to
require density distributions different from the standard uniform-outflow
assumption as indicated by the models of Monnier et al.\ (1997) and the
present models. On the other hand, Monnier et al.\ (1997) used a two-component
model with exponentially declining density distributions that is %%completely
different from the current superwind model which, in turn,
shows single density jumps but preserves
$1/r^{2}$ density courses in the various dust shell parts.
The present study matches the observations for a broad wavelength range, i.e.\
from $\sim 0.7\mu$m to $400\mu$m. Another difference concerns the choice
of other dust-shell parameters, in particular the optical constants,
the dust-shell temperature and the dust-grain sizes.
Furthermore, it has to be kept in mind that the observational data often
belong to different epochs and in the case of long-baseline interferometry
on the used position angle as well. For instance, Sudol et al.\ (1999) found
their low spatial-frequency $11\mu$m data not to be consistent with the higher
frequency data of Monnier et al.\ (1997) possibly pointing to an expanding
double-shell structure (due to the different epochs) or to the existence
of bipolar structures (due to the different position angles). They stress the
need of coeval data covering several position angles to solve this issue.
Danchi et al.\ (1999) presented follow-up observations of the intial
measurements of Monnier et al.\ (1997) giving evidence of the
dust-shell expansion at $11\mu$m. We note that it was essential for the
current superwind models to consider the Danchi et al.\ (1999) data being
coeval with our $K$-band measurements instead of taking only the former
Monnier et al.\ observations. Finally, all models of \object{NML\,Cyg}
available so far rely on the assumption of spherical symmetry whose
applicablity still has to be proven. Richards et al.\ (1996), for instance,
discussed the existence of a bipolar outflow from interferometric maser
observations. Nature is certainly much more complex than the simple models
given in the present study and great care has to be taken during
model construction concerning its reliability and unambiguousness.
Nevertheless, it can be concluded that the dust shell of \object{NML\,Cyg}
consists of a multiple-component structure with two regions
attesting the ocurrence of previous superwind phases, one in its immediate
past, i.e.\ decades ago,  and one hundreds of years ago. Thus, 
the red supergiant \object{NML\,Cyg} proves to be an important example of
episodic mass-loss phases in the evolution of massive stars along
the Red Supergiant Branch. 

\begin{acknowledgements}
The observations were made with  the SAO 6\,m telescope, operated by the
Special Astrophysical Observatory, Russia.
We thank K.\ Justtanont for providing the ISO-SWS spectral energy distribution.
The radiative-transfer calculations are based on 
the code DUSTY developed by \v{Z}.\ Ivezi\'c, M.\ Nenkova and M. Elitzur.  
This research has made use of the SIMBAD database, operated by CDS in
Strasbourg. 
\end{acknowledgements}

\begin{thebibliography}{}
%

\bibitem[1979]{BenMut79}
 Benson J.M., Mutel R.L., 1979, ApJ 233, 119

\bibitem[2000]{BobMar2000}
 Boboltz D.A., Marvel K.B., 2000, ApJ 545, L149

\bibitem[1999]{BloeEtal99}
 Bl\"ocker T., Balega Y., Hofmann K.-H., Lichtenth\"aler J., Osterbart R.,
 Weigelt G., 1999, A\&A 348, 805

\bibitem[1983]{BowEtal83}
 Bowers P.F., Johnston K.J., Spencer J.H., 1983, ApJ 274, 733

\bibitem[1999]{DanEtal99}
 Danchi W.C., Monnier J.D., Hale D.D.S., Tuthill P.G., Townes C.H., 1999,
 in Optical and IR Interferometry from Ground and Space,
 S.C. Unwin \& R.V. Stachnick (eds.), ASP Conf.\ Ser. 194, p.\ 180

\bibitem[1995]{DavPeg95}
 David P., Pegourie B, 1995, A\&A 293, 833

\bibitem[1984]{DiaEtal84}
 Diamond P.J., Norris R.P., Booth R.S., 1984, MNRAS 207, 611

\bibitem[1978]{DickEtal78}
 Dickinson D.F., Reid M.J., Morris M., Redman R., 1978, ApJ 220, L113 

\bibitem[1984]{DraLee84}
 Draine B.T., Lee H.M., 1984, ApJ 285, 89

\bibitem[1992]{DyckBen92}
 Dyck H.M., Benson J.M.., 1992, AJ 104, 377

\bibitem[1971]{DyckEtal71}
 Dyck H.M., Forbes F.F., Shawl S.J., 1971, AJ 76, 901

\bibitem[1974]{DyckEtal74}
 Dyck H.M., Lockwood G.W., Capps R.W., 1974, ApJ 189, 89

\bibitem[1998]{DyckEtal98}
 Dyck H.M., van Belle G.T., Thompson R.R., 1998, AJ 116, 981

\bibitem[1996]{DyckEtal96}
 Dyck H.M., Benson J.A., van Belle G.T., Ridgway S.T., 1996, AJ 111, 1705

\bibitem[1984]{DyckEtal84}
 Dyck H.M., Zuckerman B., Leinert C., Beckwith S., 1984, ApJ 287, 801

\bibitem[1986]{EngEtal86}
 Engels D., Schmid-Burgk J., Walmsley C.M., 1986, A\&A 167, 129

\bibitem[1988]{FixCob88}
 Fix J.D., Cobb M.L., 1988, ApJ 329, 290

\bibitem[1967]{For67}
 Forbes F.F., 1967, ApJ 147, 1226

\bibitem[1999]{GauEtal99}
    Gauger A., Balega Y., Irrgang P., Osterbart R., Weigelt G.,
    1999, A\&A 346, 505

\bibitem[1974]{GossEtal74}
 Goss W.M., Winnberg A., Habing H.J., 1974, A\&A 30, 349

\bibitem[1976]{GreSea76}
 Gregory P.C., Seaquist E.R., 1976, ApJ 204, 626

\bibitem[1982]{HabEtal82}
 Habing H.J., Goss W.M.,, Winnberg A., 1982, A\&A 108, 412

\bibitem[1972]{Hack72}
 Hackwell J.A., 1974, A\&A 21, 239

\bibitem[1974]{HarEtal74}
 Harvey P.M., Bechis K.P., Wilson W.J., Ball J.A., 1974, ApJS 27, 331

\bibitem[1997]{HegEtal97}
 Heger A., Jeannin L., Langer N., Baraffe I., 1997, A\&A 327, 224

\bibitem[1970]{HerZap70}
 Herbig G.H., Zappala R.R., 1970, ApJ 162, L15

\bibitem[1986]{HofWei86}
 Hofmann K.-H., Weigelt G., 1986, A\&A 167, L15

\bibitem[1972]{HylEtal72}
 Hyland A.R., Becklin E.E., Frogel J.A., Neugebauer G., 1972 A\&A 16, 204

\bibitem[1996]{IveEli96}
 Ivezi\'c \v{Z}., Elitzur M., 1996, MNRAS 279, 1019

\bibitem[1997]{IveEli97}
 Ivezi\'c \v{Z}., Elitzur M., 1997, MNRAS 287, 799

\bibitem[1997]{IveNenEli97}
 Ivezi\'c \v{Z}., Nenkova M. Elitzur M., 1997, User Manual for DUSTY,
 University of Kentucky
    %%%, accessible at (http://www.pa.uky.edu/\~moshe/dusty)

\bibitem[1997]{JagNie97}
  de Jager C, Nieuwenhuijzen H., 1997, MNRAS 290, L50.

\bibitem[1967]{John67}
 Johnson H.L., 1967, ApJ 149, 345

\bibitem[1968]{John68}
 Johnson H.L., 1968, ApJ 154, L125

\bibitem[1965]{JohnEtal65}
 Johnson H.L., Low F.J., Steimentz D., 1965, ApJ 142, 808

\bibitem[1996]{JustaEtal96}
 Justtanont K., de Jong T., Helmich F.P., Waters L.B.F.M., de Graauw Th.,
 Loup C., Izumiura H., Yamamura I., Beintema D.A., Lahuis F., Roelfsema P.R.,
 Valentijn E.A., 1996, A\&A 315, L217

\bibitem[1982]{KnaEtal82}
 Knapp G.R., Philips T.G., Leighton R.B., Lo K.Y., Wannier, P.G.,
 Wootten H.A., 1982, ApJ 252, 616.

\bibitem[1997]{KrueSedl97}
 Kr\"uger D., Sedlmayr E., 1997, A\&A 321, 557

\bibitem[1971]{Kru71}
 Kruzewski A., 1971, AJ 76, 576

\bibitem[1970]{Lab70}
 Labeyrie A., 1970, A\&A  6, 85

\bibitem[1998]{LanEtal98}
 Langer N., Heger A., Garcia-Segura G., 1998, Rev, Mod. Astron. 11, 57. 

\bibitem[1970]{Lee70}
 Lee T.A., 1970, ApJ 162, 217

\bibitem[1983]{LohWeiWir83}
 Lohmann A.W., Weigelt G., Wirnitzer B., 1983, Appl. Opt. 22, 4028

\bibitem[1970]{LowEtal70}
 Low F.J., Johnson H.L., Kleinman D.E., Latham A.S., Geisel S.L., 1970,
 ApJ 160, 531

\bibitem[1992]{LucEtal92}
 Lucas R., Bujarrabal V., Guilloteau, Bachiller R., Baudry A., Cernicharo J.,
 Delannoy J., Forveille T., Guelin M., Radford S.J.E., 1992, A\&A 262, 491

\bibitem[1979]{McCar79}
 McCarthy D.W. Jr., 1979, High Angular Resolution Stellar Interferometry,
 IAU Coll.\ 50, J.\ Davis \& W.J.\ Tango (eds.), 18

\bibitem[1974]{MashEtal74}
 Masheder M.R.W., Booth R.S., Davies R.D., 1974,, MNRAS 166, 561

\bibitem[1977]{MRN77}
 Mathis J.S., Rumpl W., Nordsieck K.H., 1977, ApJ 217, 425 (MRN)
    
\bibitem[1976]{MerSt76}
 Merrill K.M., Stein W.A., 1976, PASP 88, 294

\bibitem[1997]{MonEtal97}
 Monnier J.D., Bester M., Danchi W.C. Johnson M.A., Lipman E.A., Townes C.H.,
 Tuthill P.G., Geballe T.R., Nishimoto D., Kervin P.W., 1997, ApJ 481, 420

\bibitem[1998]{MonEtal98}
 Monnier J.D., Geballe T.R , Danchi W.C., 1998, ApJ 502, 833

\bibitem[1983]{MorJur83}
 Morris M., Jura M., 1983, ApJ 267, 179

\bibitem[1979]{MorEtal79}
 Morris M., Redman R., Reid M.J., Dickinson D.F., 1979, ApJ 229, 257

\bibitem[1965]{NML65}
 Neugebauer G., Martz D.E., Leighton R.B., 1965, ApJ 142, 399

\bibitem[1992]{OssEtal92}
 Ossenkopf V., Henning T., Mathis J.S., 1992, A\&A 261, 567 

\bibitem[1998]{PerEtal98}
 Perrin G., Coude du Foresto V., Ridgway S.T., Mariotti J.M., Traub W.A.,
 Carleton N.P., Lacasse M.G., 1998, A\&A 331, 619

\bibitem[1996]{RichEtal96}
 Richards A.M.S., Yates J.A., Cohen R.J., 1996, MNRAS 282, 665

\bibitem[1986]{RidgEtal86}
 Ridgway S.T., Joyce R.R., Connors D., Pipher J.L., Dainty C., 1986,
 ApJ 302, 662

\bibitem[1983]{RowRobHar83}
 Rowan-Robinson M, Harris S., 1983, MNRAS 202, 767

\bibitem[1979]{SavMat79}
 Savage B.D., Mathis J.S., 1979, ARA\&A 17, 73

\bibitem[1970]{SchwBar70}
 Schwartz P.R., Barrett A.H., 1970, ApJ 159, L123

\bibitem[1974]{SchwEtal74}
 Schwartz P.R., Harvey P.M., Barrett A.H., 1974, ApJ 187, 491

%\bibitem[1971]{Ser71}
% Serkowski K., 1971, ApJ 179, L101
%
\bibitem[1979]{SibEtal79}
 Sibille F., Chelli A., Lena P., 1979, A\&A 79. 315

\bibitem[1975]{SnyBuhl75}
 Snyder L.E., Buhl D., 1975, ApJ 197, 329

\bibitem[1985]{SopEtal85}
 Sopka R.J., Hildebrand R., Jaffe D.T., Gatley I., Roelig T., Werner M.,
 Jura M., Zuckerman B., 1985, ApJ 294, 242 

\bibitem[1975]{Str75}
 Strecker D.W., 1975, AJ 80, 451

\bibitem[1974]{StrNey74}
 Strecker D.W., Ney E.P., 1974, AJ 79, 1410

\bibitem[1999]{SudEtal99}
 Sudol J.J., Dyck H.M., Stencel R.E., Klebe DDD.I., Creech-Eakman M.J., 1999,
 AJ 117, 1609

\bibitem[1997]{SuhJon97}
 Suh K.W., Jones T.J., 1997, ApJ 479, 918

\bibitem[1994]{VanbeverEtal94}
 Vanbeveren D., van Rensbergen W., de Loore C. (eds.) , 1994, Evolution of
 Massive Stars, Kluwer, Dordrecht

\bibitem[1991]{WalEtal91}
 Walmsley C.M., Chini R., Kreysa E., Steppe H., Forveille T., Omont A.,
 1991, A\&A 248, 555

\bibitem[1977]{Wei77}
 Weigelt G., 1977, Optics Commun. 21, 55

\bibitem[1991]{Wei91}
 Weigelt G., 1991, in: Progress in Optics Vol.\,29, E.\ Wolf (ed.),
 Elsevier Science Publishers, Amsterdam,  p.\ 293

\bibitem[1970]{WilEtal70}
 Wilson W.J., Barrett A.H., Moran J.M., 1970, ApJ 160, 545

\bibitem[1997]{WinEtal97}
 Winters J.M., Fleischer A.J., Le Bertre T., Sedlmayr E., 1997, A\&A 326, 305

\end{thebibliography}
\end{document}